\documentclass[journal]{IEEEtran}
\usepackage[pdftex]{color,graphicx}
\usepackage{amssymb}
\usepackage{amsmath}

\newtheorem{theorem}{Theorem}
\newtheorem{definition}{Definition}
\newtheorem{claim}{Claim}
\newtheorem{lemma}{Lemma}

\renewcommand{\QED}{\QEDopen}

\begin{document}


\title{Low Density Lattice Codes}
\author{Naftali~Sommer,~\IEEEmembership{Senior Member,~IEEE,}
        Meir~Feder,~\IEEEmembership{Fellow,~IEEE,}
        and~Ofir~Shalvi,~\IEEEmembership{Member,~IEEE} 
\thanks{The material in this paper was presented in part in the IEEE International Symposium on Information Theory, Seattle, July 2006, and in part in the Inauguration of the UCSD Information Theory and Applications Center, San Diego, Feb. 2006.} % <-this % stops a space        
        }
\maketitle

\begin{abstract}
Low density lattice codes (LDLC) are novel lattice codes that can
be decoded efficiently and approach the capacity of the additive
white Gaussian noise (AWGN) channel. In LDLC a codeword
$\underline{x}$ is generated directly at the $n$-dimensional
Euclidean space as a linear transformation of a corresponding
integer message vector $\underline{b}$, i.e.,
$\underline{x}=\boldsymbol{G}\underline{b}$, where
$\boldsymbol{H}=\boldsymbol{G}^{-1}$ is restricted to be sparse.
The fact that $\boldsymbol{H}$ is sparse is utilized to develop a
linear-time iterative decoding scheme which attains, as
demonstrated by simulations, good error performance within $\sim
0.5$dB from capacity at block length of $n=100,000$ symbols. The
paper also discusses convergence results and implementation
considerations.
\end{abstract}

\begin{keywords}
Lattices, lattice codes, iterative decoding, LDPC.
\end{keywords}

\section{Introduction}
If we take a look at the evolution of codes for binary or finite
alphabet channels, it was first shown \cite{Shannon1} that channel
capacity can be achieved with long random codewords. Then, it was
found out \cite{Elias_linear} that capacity can be achieved via a
simpler structure of linear codes.
%in which the codewords are of the
%form $\underline{x}=\boldsymbol{G}\underline{b}$, where
%$\boldsymbol{G}$ is a binary generator matrix, $\underline{b}$ is
%the information vector and the computations use finite field arithmetic.
Then, specific families of linear codes were found that are
practical and have good minimum Hamming distance (e.g.
convolutional codes, cyclic block codes, specific cyclic codes
such as BCH and Reed-Solomon codes \cite{Blahut}). Later, capacity
achieving schemes were found, which have special structures that
allow efficient iterative decoding, such as low-density
parity-check (LDPC) codes \cite{Gallager} or turbo codes
\cite{turbo}.

If we now take a similar look at continuous alphabet codes for the
additive white Gaussian noise (AWGN) channel, it was first shown
\cite{Shannon2} that codes with long random Gaussian codewords can
achieve capacity. Later, it was shown that lattice codes can also
achieve capacity (\cite{Debuda1} %\cite{Debuda2}\cite{correct_Debuda}\cite{Lolieger}\cite{Urbanke_lattice}
 -- \cite{Zamir_Erez}).
Lattice codes are clearly the Euclidean space analogue of linear
codes.
Similarly to binary codes, we could expect that specific practical lattice codes will then be developed.
%, since the codewords are of the same form
%$\underline{x}=\boldsymbol{G}\underline{b}$, where
%$\boldsymbol{G}$ is a real matrix and $\underline{b}$ is a vector
%of integers, and the only change is real algebra instead of finite
%field algebra.
However, there was almost no further progress from
that point. Specific lattice codes that were found were based on
fixed dimensional classical lattices \cite{Sloane} or based on
algebraic error correcting codes
\cite{Calderbank_codes}\cite{Forney_codes}, but no significant
effort was made in designing lattice codes directly in the
Euclidean space or in finding specific capacity achieving lattice
codes. Practical coding schemes for the AWGN channel were based on finite alphabet codes.

In \cite{signal}, ``signal codes'' were introduced. These are
lattice codes, designed directly in the Euclidean space, where the
information sequence of integers $i_n$, $n=1,2,...$ is encoded by
convolving it with a fixed signal pattern $g_n$, $n=1,2,...d$.
Signal codes are clearly analogous to convolutional codes, and in
particular can work at the AWGN channel cutoff rate with simple
sequential decoders. In \cite{signal_journal} it is also demonstrated that signal codes can work near the AWGN channel capacity with more elaborated bi-directional decoders. Thus, signal codes provided the first step
toward finding effective lattice codes with practical decoders.

Inspired by LDPC codes and in the quest of finding practical
capacity achieving lattice codes, we propose in this work ``Low
Density Lattice Codes'' (LDLC). We show that these codes can
approach the AWGN channel capacity with iterative decoders whose
complexity is linear in block length. In recent years several
schemes were proposed for using LDPC over continuous valued
channels by either multilevel coding \cite{hou_et_al} or by
non-binary alphabet (e.g. \cite{bennatan_burshtein}). Unlike these
LDPC based schemes, in LDLC both the encoder and the channel use
the same real algebra which is natural for the continuous-valued
AWGN channel. This feature also simplifies the convergence
analysis of the iterative decoder.

The outline of this paper is as follows. Low density lattice codes
are first defined in Section \ref{def_sec}. The iterative decoder
is then presented in Section \ref{iter_dec_sec}, followed by
convergence analysis of the decoder in Section \ref{convergence}.
Then, Section \ref{code_design} describes how to choose the LDLC
code parameters, and Section \ref{implementation} discusses
implementation considerations. The computational complexity of the
decoder is then discussed in Section \ref{complexity}, followed by
a brief description of encoding and shaping in Section
\ref{enc_shape}. Simulation results are finally presented in
Section \ref{sim_res}.

\section{Basic Definitions and Properties} \label{def_sec}

\subsection{Lattices and Lattice Codes} \label{lattice_codes}
%In this subsection we shall briefly summarize basic properties of lattices and
%lattice codes (see, e.g., \cite{Sloane}).
An $n$ dimensional lattice in $\mathbb{R}^m$ is defined as the set
of all linear combinations of a given basis of $n$ linearly
independent vectors in $\mathbb{R}^m$ with integer coefficients.
The matrix $\boldsymbol{G}$, whose columns are the basis vectors, is called a
generator matrix of the lattice. Every lattice point is therefore of the form $\underline{x}=\boldsymbol{G}\underline{b}$, where $\underline{b}$ is an $n$-dimensional vector of integers.
%(we shall use $G$ also
%for denoting the lattice defined by the matrix $G$).
The Voronoi cell of a lattice point is defined as the set of all
points that are closer to this point than to any other lattice
point. The Voronoi cells of all lattice points are congruent, and
for square $\boldsymbol{G}$ the volume of the Voronoi cell is
equal to $det(\boldsymbol{G})$. In the sequel $\boldsymbol{G}$
will be used to denote both the lattice and its generator matrix.

A lattice code of dimension $n$ is defined by a (possibly shifted)
lattice $\boldsymbol{G}$ in $\mathbb{R}^m$ and a shaping region $B
\subset \mathbb{R}^m$, where the codewords are all the lattice
points that lie within the shaping region $B$. Denote the number
of these codewords by $N$. The average transmitted power (per
channel use, or per symbol) is the average energy of all
codewords, divided by the codeword length $m$. The information
rate (in bits/symbol) is $log_2(N)/m$.

When using a lattice code for the AWGN channel with power limit
$P$ and noise variance $\sigma ^2$, the maximal information rate
is limited by the capacity
$\frac{1}{2}\log_2(1+\frac{P}{\sigma^2})$. Poltyrev
\cite{Poltyrev} considered
%the case of infinite constellations for
the AWGN channel without restrictions.  If there is no power
restriction, code rate is a meaningless measure, since it can be
increased without limit. Instead, it was suggested in
\cite{Poltyrev} to use the measure of constellation density,
leading to a generalized definition of the capacity as the maximal
possible codeword density that can be recovered reliably. When
applied to lattices, the generalized capacity implies that there
exists a lattice $\boldsymbol{G}$ of high enough dimension $n$
that enables transmission with arbitrary small error
probability, if and only if $\sigma^2 < \frac{\sqrt[n]{|det(\boldsymbol{G})|^2}}{2 \pi e}$. %For
%the high SNR regime,
A lattice that achieves the generalized
capacity of the AWGN channel without restrictions, also achieves
the channel capacity of the power constrained AWGN channel, with a
properly chosen spherical shaping region (see also \cite{Zamir_Erez}).

In the rest of this work we shall concentrate on the lattice design
and the lattice decoding algorithm, and not on the shaping region
or shaping algorithms. We shall use lattices with $det(\boldsymbol{G})=1$,
where analysis and simulations will be carried for the AWGN
channel without restrictions. A capacity achieving lattice will
have small error probability for noise variance $\sigma^2$ which
is close to the theoretical limit $\frac{1}{2 \pi e}$.
%According
%to Poltyrev's results, such a lattice will also be capacity
%approaching when used for the AWGN channel with power constraint.

\subsection{Syndrome and Parity Check Matrix for Lattice Codes}
A binary $(n, k)$ error correcting code is defined by its $n \times k$
binary generator matrix $\boldsymbol{G}$. A binary information vector $\underline{b}$ with dimension $k$ is encoded by $\underline{x}=\boldsymbol{G}\underline{b}$, where calculations are performed in the finite field GF(2). The parity check matrix $\boldsymbol{H}$ is an $(n-k) \times n$ matrix such that $\underline{x}$ is a codeword if and only if $\boldsymbol{H}\underline{x}=\underline{0}$. The input to the decoder is the noisy codeword $\underline{y}=\underline{x}+\underline{e}$, where $\underline{e}$ is the error sequence and addition is done in the finite field. The decoder typically starts by calculating the syndrome $\underline{s}=\boldsymbol{H}\underline{y}=\boldsymbol{H}(\underline{x}+\underline{e})=\boldsymbol{H}\underline{e}$ which depends only on the noise sequence and not on the transmitted codeword.

We would now like to extend the definitions of the parity check matrix and the syndrome to lattice codes.
An $n$-dimensional lattice code is defined by its $n \times n$ lattice generator matrix $\boldsymbol{G}$
%and by a shaping region, as explained above
(throughout this paper we assume that $\boldsymbol{G}$ is square,
but the results are easily extended to the non-square case).
%Encoding of an integer valued information vector $\underline{i}$ is performed in two steps. First (the shaping step) the integer vector $\underline{i}$ is mapped to a (possibly different) integer vector $\underline{b}$. Then (the encoding step) the codeword is generated by $\underline{x}=\boldsymbol{G}\underline{b}$. The shaping step ensures that the resulting codeword will be inside the shaping region.
Every codeword is of the form
$\underline{x}=\boldsymbol{G}\underline{b}$, where $\underline{b}$
is a vector of integers. Therefore,
$\boldsymbol{G}^{-1}\underline{x}$ is a vector of integers for
every codeword $\underline{x}$. We define the parity check matrix
for the lattice code as $\boldsymbol{H} \stackrel{\Delta}{=}\boldsymbol{G}^{-1}$. Given a
noisy codeword $\underline{y}=\underline{x}+\underline{w}$ (where
$w$ is the additive noise vector, e.g. AWGN, added by real
arithmetic), we can then define the syndrome as $\underline{s}
\stackrel{\Delta}{=}frac\{\boldsymbol{H}\underline{y}\}$, where
$frac\{x\}$ is the fractional part of $x$, defined as $frac\{x\}=
x - \left\lfloor x \right\rceil$, where $\left\lfloor x
\right\rceil$ denotes the nearest integer to $x$. The syndrome
$\underline{s}$ will be zero if and only if $\underline{y}$ is a
lattice point, since $\boldsymbol{H}\underline{y}$ will then be a
vector of integers with zero fractional part. For a noisy
codeword, the syndrome will equal
$\underline{s}=frac\{\boldsymbol{H}\underline{y}\}=frac\{\boldsymbol{H}(\underline{x}+\underline{w})\}=frac\{\boldsymbol{H}\underline{w}\}$
and therefore will depend only on the noise sequence and not on
the transmitted codeword, as desired.

Note that the above definitions of the syndrome and parity check matrix for lattice codes are consistent with the definitions of the dual lattice and the dual code\cite{Sloane}: the dual lattice of a lattice $\boldsymbol{G}$ is defined as the lattice with generator matrix $\boldsymbol{H}=\boldsymbol{G}^{-1}$, where for binary codes, the dual code of $\boldsymbol{G}$ is defined as the code whose generator matrix is $\boldsymbol{H}$, the parity check matrix of $\boldsymbol{G}$.

\subsection{Low Density Lattice Codes}
We shall now turn to the definition of the codes proposed in this
paper - low density lattice codes (LDLC).
%These codes are inspired
%by LDPC codes, which are defined as linear block codes for which
%the parity check matrix is sparse. LDLC are defined in the same
%manner, according to the above definition of a parity check matrix
%for a lattice code.

\begin{definition}[LDLC] \label{ldlc_def}
An $n$ dimensional LDLC is an $n$-dimensional lattice code with a
non-singular lattice generator matrix $\boldsymbol{G}$ satisfying
$|det(\boldsymbol{G})|=1$, for which the parity check matrix
$\boldsymbol{H}=\boldsymbol{G}^{-1}$ is sparse. The $i$'th row
degree $r_i$, $i=1, 2, ...n$ is defined as the number of nonzero
elements in row $i$ of $\boldsymbol{H}$, and the $i$'th column
degree $c_i$, $i=1, 2, ...n$ is defined as the number of nonzero
elements in column $i$ of $\boldsymbol{H}$.
\end{definition}

Note that in binary LDPC codes, the code is completely defined by the locations of the nonzero
elements of $\boldsymbol{H}$.
In LDLC there is another degree of freedom since we also have to choose the \emph{values}
of the nonzero elements of $\boldsymbol{H}$.

\begin{definition}[regular LDLC]
An $n$ dimensional LDLC is regular if all the row degrees and column degrees
of the parity check matrix are equal to a common degree $d$. %, i.e. $r_i=d$ and $c_i=d$ $\forall i=1,2,...n$.
\end{definition}

\begin{definition}[magic square LDLC] \label{magic_def}
An $n$ dimensional regular LDLC with degree $d$ is called ``magic
square LDLC'' if every row and column of the parity check matrix
$\boldsymbol{H}$ has the same $d$ nonzero values, except for a
possible change of order and random signs. The sorted sequence of these $d$ values $h_1 \geq h_2 \geq ... \geq h_d > 0$ will be referred to as the generating sequence of the magic square LDLC.
\end{definition}

For example, the matrix
\begin{displaymath}
\mathbf{\boldsymbol{H}} =
\left(\begin{array}{cccccc}
0 & -0.8 & 0 & -0.5 & 1 & 0 \\
0.8 & 0 & 0 & 1 & 0 & -0.5 \\
0 & 0.5 & 1 & 0 & 0.8 & 0 \\
0 & 0 & -0.5 & -0.8 & 0 & 1 \\
1 & 0 & 0 & 0 & 0.5 & 0.8 \\
0.5 & -1 & -0.8 & 0 & 0 & 0
\end{array} \right)
\end{displaymath}
is a parity check matrix of a magic square LDLC with lattice
dimension $n=6$, degree $d=3$ and generating sequence $\{1,
0.8, 0.5\}$. This $\boldsymbol{H}$ should be further normalized by the constant
$\sqrt[n]{|det(\boldsymbol{H})|}$ in order to have $|det(\boldsymbol{H})|=|det(\boldsymbol{G})|=1$, as
required by Definition \ref{ldlc_def}.

\begin{figure}
\centering
%\vspace {-0.0cm}
\includegraphics[height=4.2in]{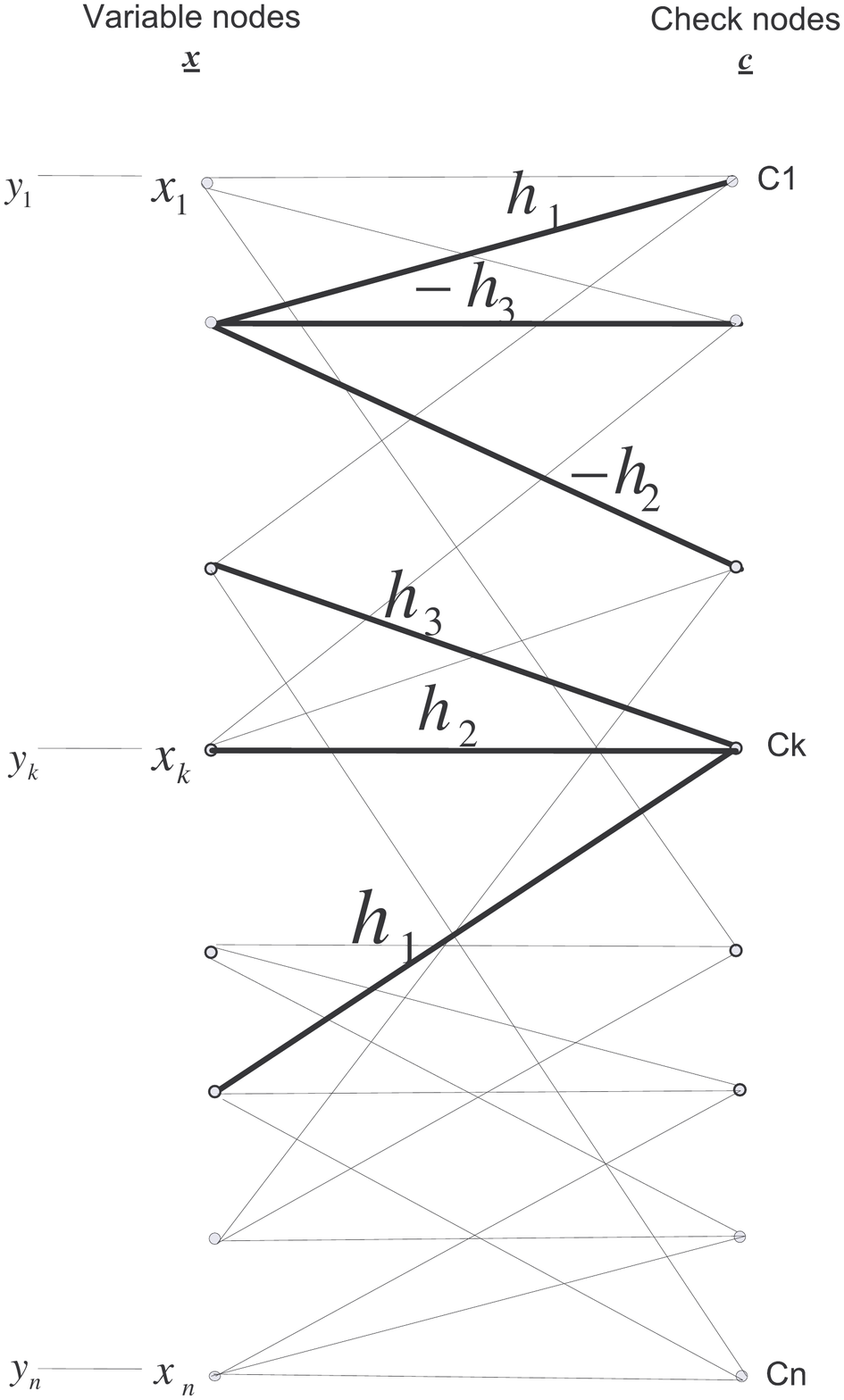}  %, width=2.5in
%\vspace {-0.0cm}
\caption{The bi-partite graph of an LDLC}
%\vspace{-0.0cm}
\label{bi_partite_fig}
\end{figure}

The bipartite graph of an LDLC is defined similarly to LDPC codes: it is a graph
with variable nodes at one side and check nodes at the other side.
Each variable node corresponds to a single element of the codeword
$\underline{x}=\boldsymbol{G}\underline{b}$. Each check node corresponds to
a check equation (a row of $\boldsymbol{H}$). A check equation is of the form $\sum_k h_k x_{i_k} = integer$, where $i_k$ denotes the locations of the nonzero elements at the appropriate row of $\boldsymbol{H}$, $h_k$ are the values of $\boldsymbol{H}$ at these locations and the integer at the right hand side is unknown. An edge connects check node $i$ and variable
node $j$ if and only if $H_{i,j} \neq 0$. This edge is assigned the value $H_{i,j}$. Figure \ref{bi_partite_fig} illustrates the bi-partite graph of a magic square LDLC with degree 3. In the figure, every variable node $x_k$ is also associated with its noisy channel observation $y_k$.

Finally, a $k$-loop is defined as a loop in the bipartite graph that consists of $k$ edges. A bipartite graph, in general, can only contain loops with even length. Also, a 2-loop, which consists of two parallel edges that originate from the same variable node to the same check node, is not possible by the definition of the graph. However, longer loops are certainly possible. For example, a 4-loop exists when two variable nodes are both connected to the same pair of check nodes.

\section{Iterative Decoding for the AWGN Channel} \label{iter_dec_sec}

%Denote the LDLC lattice by $G$ and the shaping region by $B$.
%We shall now turn to the LDLC decoding problem for the AWGN channel. Denote the $n$-dimensional lattice generated by $\boldsymbol{G}$ by $\Lambda$ and the shaping region by $B$.
Assume that the codeword $\underline{x}=\boldsymbol{G}\underline{b}$ was transmitted,
where $\underline{b}$ is a vector of integers.
We observe the noisy codeword $\underline{y}=\underline{x}+\underline{w}$,
where $\underline{w}$ is a vector of i.i.d Gaussian noise samples with common variance $\sigma^2$,
and we need to estimate the integer valued vector $\underline{b}$.
The maximum likelihood (ML) estimator is then %$\hat{\underline{b}}=\arg \mathop{\max }\limits_{\underline{b}} Pr\{\underline{b}|\underline{y}\}$. For the AWGN channel, it is well known that the ML estimator is
$\hat{\underline{b}}=\arg \mathop{\min }\limits_{\underline{b}}||\underline{y}-\boldsymbol{G}\underline{b}||^2$.

Our decoder will not estimate directly the integer vector $\underline{b}$. Instead, it will estimate the probability density function (PDF) of the codeword vector $\underline{x}$.
Furthermore, instead of calculating the $n$-dimensional PDF of the whole vector $\underline{x}$, we shall calculate the $n$ one-dimensional PDF's for each of the components $x_k$ of this vector (conditioned on the whole observation vector $\underline{y}$). In appendix \ref{app_exact_pdf} it is shown that $f_{x_k|\underline{y}}(x_k|\underline{y})$ is a weighted sum of Dirac delta functions:
\begin{eqnarray} \label{exact_solution}
f_{x_k|\underline{y}}(x_k|\underline{y}) %= \int\int_{x_i, i \neq k}\cdots\int f_{\underline{x}|\underline{y}}(\underline{x}|\underline{y}) d x_1 d x_2 \cdots d x_{k-1} d x_{k+1} \cdots d x_n =
%\nonumber\\
= C \cdot \sum_{\underline{l} \in \boldsymbol{G} \cap B} \delta (x_k - l_k) \cdot e^{-d^2(\underline{l}, \underline{y})/2\sigma^2}
\end{eqnarray}
where $\underline{l}$ is a lattice point (vector), $l_k$ is its
$k$-th component, $C$ is a constant independent of $x_k$ and
$d(\underline{l},\underline{y})$ is the Euclidean distance between
$\underline{l}$ and $\underline{y}$.
%We would like to choose the delta function with largest amplitude in order to get the ML estimation for $x_k$. However, This will not necessarily give the $k'th$ component of the ML solution for $\underline{x}$. The reason is that several lattice points may have the same value for $x_k$, so their corresponding delta functions may add together and have larger amplitude than the delta function that corresponds to the value of $x_k$ for the ML solution.  Therefore, estimating each $x_k$ separately is only an approximate ML solution. In any case,
Direct evaluation of (\ref{exact_solution}) is not practical, so
our decoder will try to estimate
$f_{x_k|\underline{y}}(x_k|\underline{y})$ (or at least
approximate it) in an iterative manner.
%\subsection{Gallager's approach} \label{sec_simp}
% We shall derive the iterative decoder by following the guidelines of Gallager's derivation for LDPC codes \cite{Gallager}.
%To simplify the derivation, Gallager used the following trick.

Our decoder will decode to the infinite lattice, thus ignoring the shaping region boundaries. This approximate decoding method is no longer exact maximum likelihood decoding, and is usually denoted ``lattice decoding'' \cite{Zamir_Erez}.

The calculation of
$f_{x_k|\underline{y}}(x_k|\underline{y})$ is
involved since the components $x_k$ are not independent random
variables (RV's), because $\underline{x}$ is restricted to be a
lattice point. Following \cite{Gallager} we use a ``trick'' - we
assume that the $x_k$'s are independent,
%i.i.d with a properly chosen PDF
%$f^{(i.i.d)}_{x_k}(x_k)$. This PDF is chosen to be the marginal
%distribution of a vector distribution which is
%$f^{(i.i.d)}_{\underline{x}}(\underline{x})$,
 %where
%$f^{(i.i.d)}_{\underline{x}}(\underline{x}) = const$.
%$\forall= \underline{x} \in  G \cap B$.
%For example, $f^{(i.i.d)}_{\underline{x}}(\underline{x})$ can be
%chosen to be
%uniformly distributed over the shaping region. In
%addition, we
 but add a condition that assures that $\underline{x}$ is a lattice point.
% only lattice points have nonzero probability.
 Specifically, define $\underline{s}
\stackrel{\Delta}{=}\boldsymbol{H} \cdot \underline{x}$.
%Define further the set of integer valued vectors $i_B$ as $i_B =
%\{\underline{i} | \underline{i} \in \mathbb{Z}^n,
%\boldsymbol{G}\underline{i} \in B\}$.
Restricting $\underline{x}$
to be a lattice point %inside the shaping region $B$
is equivalent
 to restricting $\underline{s} \in \mathbb{Z}^n$. %$\underline{s} \in i_B$.
 Therefore, instead of calculating
$f_{x_k|\underline{y}}(x_k|\underline{y})$
under the assumption that $\underline{x}$ is a lattice point, %in $B$,
we can calculate $f_{x_k|\underline{y}}(x_k|\underline{y},\underline{s}\in \mathbb{Z}^n)$ and assume
that the $x_k$ are independent and identically distributed (i.i.d) with a continuous PDF (that does not include Dirac delta functions). % Appendix \ref{trick_example} illustrates this technique by calculating $f_{\underline{x}|\underline{y}}(\underline{x}|\underline{y})$ in an alternative way to the direct calculation done in appendix \ref{app_exact_pdf}.
It still remains to set $f_{x_k}(x_k)$, the PDF of $x_k$.
Under the i.i.d assumption, the PDF of the codeword $\underline{x}$ is $f_{\underline{x}}(\underline{x})=\prod_{k=1}^n f_{x_k}(x_k)$. As shown in Appendix \ref{trick_pdf}, the value of $f_{\underline{x}}(\underline{x})$ is not important at values of $\underline{x}$ which are not lattice points, but at a lattice point it should be proportional to the probability of using this lattice point. Since we assume that all lattice points are used equally likely, $f_{\underline{x}}(\underline{x})$ must have the same value at all lattice points. A reasonable choice for $f_{x_k}(x_k)$ is then to use a uniform distribution such that $\underline{x}$ will be uniformly distributed in an $n$-dimensional cube. For an exact ML decoder (that takes into account the boundaries of the shaping region), it is enough to choose the range of $f_{x_k}(x_k)$ such that this cube will contain the shaping region. For our decoder, that performs lattice decoding, we should set the range of $f_{x_k}(x_k)$ large enough such that the resulting cube will include all the lattice points which are likely to be decoded. The derivation of the iterative decoder shows that this range can be set as large as needed without affecting the complexity of the decoder.

\begin{figure}
\centering
%\vspace {-0.3cm}
\includegraphics[height=1.5in]{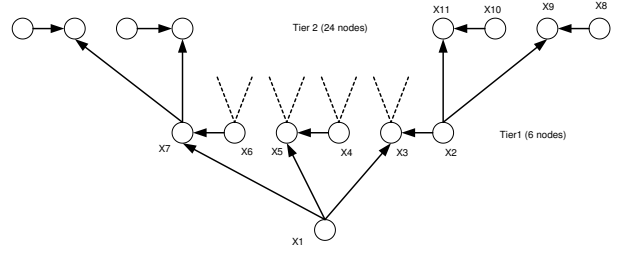}  %, width=2.5in
%\vspace {-0.5cm}
\caption{Tier diagram}
%\vspace{-0.6cm}
\label{tier_diagram}
\end{figure}

The derivation in \cite{Gallager} further imposed the tree assumption. In order to understand the tree assumption, it is useful to define the tier diagram, which is shown in %, which can be demonstrated using the tier diagram.
Figure \ref{tier_diagram} for a regular LDLC with degree 3. Each vertical line corresponds to a check equation. The tier 1 nodes of $x_1$ are all the elements $x_k$ that take place in a check equation with $x_1$. %In the example shown, these equations are of the form:
%\begin{eqnarray}
%h_{11} \cdot x_1 + h_{12} \cdot x_2 + h_{13} \cdot x_3 = integer
%\end{eqnarray}
%\begin{eqnarray}
%h_{21} \cdot x_1 + h_{22} \cdot x_4 + h_{23} \cdot x_5 = integer
%\end{eqnarray}
%\begin{eqnarray}
%h_{31} \cdot x_1 + h_{32} \cdot x_6 + h_{33} \cdot x_7 = integer
%\end{eqnarray}
%where $h_{ij}$ denotes the values of the elements of the parity check matrix $H$ in the appropriate nonzero locations.
The tier 2 nodes of $x_1$ are all the elements that take place in check equations with the tier 1 elements of $x_1$, and so on. The tree assumption assumes that all the tree elements are distinct (i.e. no element appears in different tiers or twice in the same tier). This assumption simplifies the derivation, but in general, does not hold in practice, so our iterative algorithm is not guaranteed to converge to the exact solution (\ref{exact_solution}) (see Section \ref{convergence}). %The convergence properties of the algorithm will be treated in section \ref{convergence}.%The tree assumption certainly simplifies the calculations. However, It is obviously not true for an LDLC with finite dimension, since the elements of $\underline{x}$ must repeat after a finite number of tiers.
The detailed derivation of the iterative decoder (using the above ``trick'' and the tree assumption) is given in Appendix \ref{iter_deriv}. In Section \ref{decoder} below we present the final resulting
algorithm. This iterative algorithm can also be explained by
intuitive arguments, described after the algorithm specification.

\subsection{The Iterative Decoding Algorithm} \label{decoder}

The iterative algorithm is most conveniently represented by using
a message passing scheme over the bipartite graph of the code,
similarly to LDPC codes. The basic difference is that in LDPC
codes the messages are scalar values (e.g. the log likelihood ratio of
a bit), where for LDLC the messages are real functions over the
interval $(-\infty,\infty)$. As in LDPC, in each iteration the
check nodes send messages to the variable nodes along the edges of
the bipartite graph and vice versa. The messages sent by the check
nodes are periodic extensions of PDF's. The messages sent by the
variable nodes are PDF's.

%\begin{description}
\underline{LDLC iterative decoding algorithm:}

Denote the variable nodes by $x_1,x_2,...,x_n$ and the check nodes
by $c_1,c_2,...c_n$.
\begin{itemize}
\item \emph{Initialization}: each variable node $x_k$ sends to all its check nodes the message $
%\begin{eqnarray} \label{init_eq}
f_k^{(0)}(x) = \frac{1}{\sqrt{2 \pi \sigma^2}} e^{-\frac{(y_k-x)^2}{2\sigma^2}}
%\end{eqnarray}
$.
\item \emph{Basic iteration - check node message}: Each check node sends a (different) message to each of the variable nodes that are connected to it. For a specific check node denote (without loss of generality) the appropriate check equation by
$%\begin{eqnarray}
\sum_{l=1}^{r}h_l x_{m_l} = integer
$%\end{eqnarray}
, where $x_{m_l}$, $l=1,2...r$ are the variable nodes that are connected to this check node (and $r$ is the appropriate row degree of $\boldsymbol{H}$). Denote by $f_l(x)$, $l=1,2...r$, the message that was sent to this check node by variable node $x_{m_l}$ in the previous half-iteration.
The message that the check node transmits back to variable node $x_{m_j}$ is calculated in three basic steps.
\begin{enumerate}
\item \emph{The convolution step} - all messages, except $f_j(x)$, are convolved (after expanding each $f_l(x)$ by $h_l$):
\begin{eqnarray} \label{iter_conv}
\tilde{p}_j(x) =
f_1\left(\frac{x}{h_1}\right) \circledast
%f_2\left(\frac{x}{h_2}\right) \circledast \cdots
%\nonumber\\
\cdots f_{j-1}\left(\frac{x}{h_{j-1}}\right)  \circledast
\nonumber\\
\circledast f_{j+1}\left(\frac{x}{h_{j+1}}\right) \circledast \cdots
%\nonumber\\
\cdots \circledast f_r\left(\frac{x}{h_r}\right)
\end{eqnarray}
\item \emph{The stretching step} -
The result is stretched by $(-h_j)$ to
$%\begin{eqnarray} \label{iter_stretch}
p_j(x) = \tilde{p}_j(-h_j x)
$%\end{eqnarray}
\item \emph{The periodic extension step} -
The result is extended to a periodic function with period $1/|h_j|$:
\begin{eqnarray}  \label{check_message}
Q_j(x) = \sum_{i=-\infty}^{\infty}p_j\left(x-\frac{i}{h_j}\right)
\end{eqnarray}
\end{enumerate}
The function $Q_j(x)$ is the message that is finally sent to variable node $x_{m_j}$.
\item \emph{Basic iteration - variable node message:} Each variable node sends a (different) message to each of the check nodes that are connected to it. For a specific variable node $x_k$, assume that it is connected to check nodes $c_{m_1},c_{m_2},...c_{m_e}$, where $e$ is the appropriate column degree of $\boldsymbol{H}$. Denote by $Q_l(x)$, $l=1,2,...e$, the message that was sent from check node $c_{m_l}$ to this variable node in the previous half-iteration. The message that is sent back to check node $c_{m_j}$ is calculated in two basic steps:
\begin{enumerate}
\item \emph{The product step}:
$%\begin{eqnarray} \label{var_prod_step}
\tilde{f}_j(x) = e^{-\frac{(y_k-x)^2}{2\sigma^2}} \prod_{\substack{l=1 \\ l \neq j}}^{e} Q_l(x)
$%\end{eqnarray}
\item \emph{The normalization step}: \label{var_norm_step}
$%\begin{eqnarray} \label{var_message}
f_j(x) = \frac{\tilde{f}_j(x)}{\int_{-\infty}^{\infty}\tilde{f}_j(x) dx}
$%\end{eqnarray}
\end{enumerate}
This basic iteration is repeated for the desired number of iterations.
\item \emph{Final decision:} After finishing the iterations, we want to estimate
the integer information vector $\underline{b}$. First, we estimate the final PDF's
of the codeword elements $x_k$, $k=1,2,...n$, by calculating the variable node
messages at the last iteration without omitting any check node message in the product
step:
$\tilde{f}^{(k)}_{final}(x) = e^{-\frac{(y_k-x)^2}{2\sigma^2}} \prod_{l=1}^{e} Q_l(x)$.
Then, we estimate each $x_k$ by finding the peak of its PDF:
$\hat{x_k} = arg \max_x \tilde{f}^{(k)}_{final}(x)$.
Finally, we estimate $\underline{b}$ as
$\underline{\hat{b}}=\left\lfloor \boldsymbol{H}\underline{\hat{x}}\right\rceil$.
%Each check node estimates the relevant integer, by calculating the convolution
%$\tilde{p}(x)$ as in (\ref{iter_conv}), but this time without omitting any PDF.
%(i.e. all the received variable node messages are convolved).
%Then, the integer $b_m$ is determined by
%$\hat{b_m} = arg \max_{j \in \mathbb{Z}} \tilde{p}(j)$.
%, as described above (the check node that corresponds to the $k$'th row of $H$
\end{itemize}
%
%\subsection{Qualitative explanation}
\label{qual_exp}
\begin{figure}
\centering
%\vspace {-0.3cm}
\includegraphics[width=3.5in]{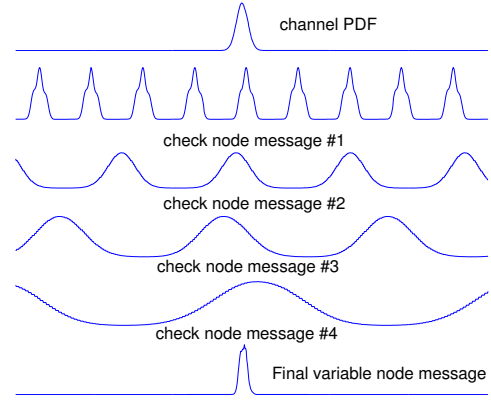}  %, width=2.5in [width=3.0in, height=1.5in]
%\vspace {-0.7cm}
\caption{Signals at variable node}
%\vspace{-0.6cm}
\label{check_waveforms}
\end{figure}

The operation of the iterative algorithm can be intuitively explained as follows.
The check node operation is equivalent to calculating the PDF of $x_{m_j}$ from the PDF's of $x_{m_i}$, $i=1,2,...,j-1, j+1,...r$, given that $\sum_{l=1}^{r}h_l x_{m_l} = integer$, and assuming that $x_{m_i}$ are independent. Extracting $x_{m_j}$ from the check equation, we get $x_{m_j} = \frac{1}{h_j}(integer - \sum_{\substack{l=1 \\ l \neq j}}^{r} h_l x_{m_l})$. Since the PDF of a sum of independent random variables is the convolution of the corresponding PDF's, equation (\ref{iter_conv}) and the stretching step that follows it simply calculate the PDF of $x_{m_j}$, assuming that the integer at the right hand side of the check equation is zero. The result is then periodically extended such that a properly shifted copy exists for every possible value of this (unknown) integer.
The variable node gets such a message from all the check equations that involve the corresponding variable. The check node messages and the channel PDF are treated as independent sources of information on the variable, so they are multiplied all together.

Note that the periodic extension step at the check nodes is equivalent to a convolution with an infinite impulse train. With this observation, the operation of the variable nodes is completely analogous to that of the check nodes: the variable nodes multiply the incoming messages by the channel PDF, where the check nodes convolve the incoming messages with an impulse train, which can be regarded as a generalized ``integer PDF''.

In the above formulation, the integer information vector $\underline{b}$ is recovered from the PDF's of the codeword elements $x_k$. An alternative approach is to calculate the PDF of each integer element $b_m$ directly as the PDF of the left hand side of the appropriate check equation. Using the tree assumption, this can be done by simply calculating the convolution
$\tilde{p}(x)$ as in (\ref{iter_conv}), but this time without omitting any PDF, 
i.e. all the received variable node messages are convolved.
Then, the integer $b_m$ is determined by
$\hat{b}_m = arg \max_{j \in \mathbb{Z}} \tilde{p}(j)$.

Figure \ref{check_waveforms} shows an example for a regular LDLC with degree $d=5$. The figure shows all the signals that are involved in generating a variable node message for a certain variable node. The top signal is the channel Gaussian, centered around the noisy observation of the variable. The next 4 signals are the periodically extended PDF's that arrived from the check nodes, and the bottom signal is the product of all the 5 signals.
It can be seen that each periodic
signal has a different period, according to the relevant
coefficient of $\boldsymbol{H}$. Also, the signals with larger period have
larger variance. This diversity resolves all the ambiguities such
that the multiplication result (bottom plot) remains with a single
peak. We expect the iterative algorithm to converge to a solution
where a single peak will remain at each PDF, located at the
desired value and narrow enough to estimate the information.

\section {Convergence}
\label{convergence}
\subsection{The Gaussian Mixture Model}
Interestingly, for LDLC we can come up with a convergence analysis
that in many respects is more specific than the similar analysis
for LDPC.

We start by introducing basic claims about Gaussian PDF's. Denote
$G_{m,V}(x) = \frac{1}{\sqrt{2 \pi V}} e^{-\frac{(x-m)^2}{2V}}$.
\begin{claim}[convolution of Gaussians] \label{conv_claim}
The convolution of $n$ Gaussians with mean values $m_1,m_2,...,m_n$ and variances $V_1,V_2,...,V_n$, respectively, is a Gaussian with mean $m_1+m_2+...+m_n$ and variance $V_1+V_2+...+V_n$.
\end{claim}

\begin{proof}
See \cite{Papoulis}.
\end{proof}

\begin{claim}[product of $n$ Gaussians] \label{product_claim}
Let $G_{m_1,V_1}(x)$, $G_{m_2,V_2}(x)$,...,$G_{m_n,V_n}(x)$ be $n$ Gaussians with mean values $m_1,m_2,...,m_n$ and variances $V_1,V_2,...,V_n$ respectively. Then, the product of these Gaussians is a scaled Gaussian: $\prod_{i=1}^n G_{m_i,V_i}(x) = \hat{A} \cdot G_{\hat{m},\hat{V}}(x)$, where $\frac{1}{\hat{V}} = \sum_{i=1}^n \frac{1}{V_i}$, $\hat{m} = \frac{\sum_{i=1}^n m_i V_i^{-1}}{\sum_{i=1}^n V_i^{-1}}$, and $\hat{A} = \frac{1}{\sqrt{(2 \pi)^{n-1} \hat{V}^{-1} \prod_{k=1}^n V_k}} \cdot e^{-\frac{\hat{V}}{2} \sum_{i=1}^n \sum_{j=i+1}^n \frac{(m_i-m_j)^2}{V_i \cdot V_j}}$.
%An alternative formulation for the scaling factor is $\hat{A} = \frac{1}{\sqrt{(2 \pi)^{n-1} \hat{V}^{-1} \prod_{k=1}^n V_k}} \cdot e^{\frac{1}{2} \left(\frac{\hat{m}^2}{\hat{V}} - \sum_{i=1}^n \frac{m_i^2}{V_i}\right)}$
\end{claim}

\begin{proof}
By straightforward mathematical manipulations.
\end{proof}

The reason that we are interested in the properties of Gaussian PDF's lies in the following lemma.

\begin{lemma}
Each message that is exchanged between the check nodes and variable nodes in the LDLC decoding algorithm (i.e. $Q_j(x)$ and $f_j(x)$), at every iteration, can be expressed as a Gaussian mixture of the form
$M(x) = \sum_{j=1}^{\infty}A_j G_{m_j,V_j}(x)$.
\end{lemma}
\begin{proof}
By induction: The initial messages are Gaussians, and the basic operations of the iterative decoder preserve the Gaussian mixture nature of Gaussian mixture inputs (convolution and multiplication preserve the Gaussian nature according to claims \ref{conv_claim} and \ref{product_claim}, stretching, expanding  and shifting preserve it by the definition of a Gaussian, and periodic extension transforms a single Gaussian to a mixture and a mixture to a mixture).
\end{proof}
Convergence analysis should therefore analyze the convergence of the variances, mean values and amplitudes of the Gaussians in each mixture.

\subsection{Convergence of the Variances} \label{var_convergence}
We shall now analyze the behavior of the variances, and start with the following lemma.
\begin{lemma}
For both variable node messages and check node messages, all the Gaussians that take place in the same mixture have the same variance.
\end{lemma}
\begin{proof}
By induction. The initial variable node messages are single element mixtures so the claim obviously holds. Assume now that all the variable node messages at iteration $t$ are mixtures where all the Gaussians that take place in the same mixture have the same variance. In the convolution step (\ref{iter_conv}), each variable node message is first expanded. All Gaussians in the expanded mixture will still have the same variance, since the whole mixture is expanded together. Then, $d-1$ expanded Gaussian mixtures are convolved. In the resulting mixture, each Gaussian will be the result of convolving $d-1$ single Gaussians, one from each mixture. According to claim \ref{conv_claim}, all the Gaussians in the convolution result will have the same variance, which will equal the sum of the $d-1$ variances of the expanded messages. The stretching and periodic extension (\ref{check_message}) do not change the equal variance property, so it holds for the final check node messages. The variable nodes multiply $d-1$ check node messages. Each Gaussian in the resulting mixture is a product of $d-1$ single Gaussians, one from each mixture, and the channel noise Gaussian. According to claim \ref{product_claim}, they will all have the same variance. The final normalization step does not change the variances so the equal variance property is kept for the final variable node messages at iteration $t+1$.
\end{proof}

Until this point we did not impose any restrictions on the LDLC. From now on, we shall restrict ourselves to magic square regular LDLC (see Definition \ref{magic_def}). %Recall (see definition \ref{magic_def}) that for such codes, the same $d$ nonzero values (except for random signs and order) appear on each row or column of $H$, and their sorted sequence is defined as the generating sequence of code. Denote the values of this sequence by $h_1 \geq h_2 \geq ...\geq h_d > 0$.
The basic iterative equations that relate the variances at iteration $t+1$ to the variances at iteration $t$ are summarized in the following two lemmas.
\begin{lemma} \label{same_val_same_var}
For magic square LDLC, variable node messages that are sent at the same iteration along edges with the same absolute value have the same variance.
\end{lemma}
\begin{proof}
See Appendix \ref{var_app}.
\end{proof}
\begin{lemma} \label{var_recur_lemma}
For magic square LDLC with degree $d$, denote the variance of the messages that are sent at iteration $t$ along edges with weight $\pm h_l$ by $V^{(t)}_{l}$. The variance values $V^{(t)}_1, V^{(t)}_2,...,V^{(t)}_d$ obey the following recursion:
\begin{eqnarray} \label{var_recur}
\frac{1}{V^{(t+1)}_i} = \frac{1}{\sigma^2} + \sum_{\substack{m=1 \\ m \neq i}}^d \frac{h_m^2}{\sum_{\substack{j=1 \\ j \neq m}}^d h_j^2 V^{(t)}_j}
\end{eqnarray}
for $i=1,2,...d$, with initial conditions $V^{(0)}_1 = V^{(0)}_2   =... = V^{(0)}_d = \sigma^2$.
\end{lemma}
\begin{proof}
See Appendix \ref{var_app}.
\end{proof}

For illustration, the recursion for the case $d=3$ is:
\begin{eqnarray} \label{var_recur_3}
\frac{1}{V^{(t+1)}_1} = \frac{h_2^2}{h_1^2 V^{(t)}_1 + h_3^2 V^{(t)}_3} + \frac{h_3^2}{h_1^2 V^{(t)}_1 + h_2^2 V^{(t)}_2} + \frac{1}{\sigma^2}
\end{eqnarray}
\begin{displaymath}
\frac{1}{V^{(t+1)}_2} = \frac{h_1^2}{h_2^2 V^{(t)}_2 + h_3^2 V^{(t)}_3} + \frac{h_3^2}{h_1^2 V^{(t)}_1 + h_2^2 V^{(t)}_2} + \frac{1}{\sigma^2}
\end{displaymath}
\begin{displaymath}
\frac{1}{V^{(t+1)}_3} = \frac{h_1^2}{h_2^2 V^{(t)}_2 + h_3^2 V^{(t)}_3} + \frac{h_2^2}{h_1^2 V^{(t)}_1 + h_3^2 V^{(t)}_3} + \frac{1}{\sigma^2}
\end{displaymath}

%Consider first the simple case of $h_1=h_2=...=h_d$. From (\ref{var_recur}), (\ref{var_recur_3}) it can be seen that if $V^{(0)}_i$ is independent of $i$, then $V^{(t)}_i$ is independent of $i$ for every $t>0$, since all the vector elements will follow the same recursive equations. Substituting this result in the first equation, we get the single variable recursion
%$\frac{1}{V^{(t+1)}_i} = \frac{1}{V^{(t)}_i} + \frac{1}{\sigma^2}$, with initialization
%$V^{(0)}_i = \sigma^2$.
%This recursion is easily solved to get
%$\frac{1}{V^{(t)}_i} = \frac{t}{\sigma^2}$
%, or
%$V^{(t)}_i = \frac{\sigma^2}{t}$.
%It can be seen that the variances approach zero as $t \rightarrow \infty$. This means that the Gaussians will approach Dirac impulses. Recall that we hope our iterative algorithm to approximate $f_{x_k|\underline{y}}(x_k|\underline{y})$, which was calculated in (\ref{exact_solution}) and was shown to be a sum of Dirac impulses. Therefore, we should not be surprised that the variances approach zero.

The lemmas above are used to prove the following theorem regarding the convergence of the variances.

\begin{theorem} \label{var_behavior}
For a magic square LDLC with degree $d$ and generating sequence $h_1 \geq h_2 \geq ...\geq h_d > 0$, define $\alpha \stackrel{\Delta}{=} \frac{\sum_{i=2}^d h_i^2}{h_1^2}$. Assume that
$\alpha < 1$. Then:
\begin{enumerate}
\item The first variance approaches a constant value of $\sigma^2(1-\alpha)$, where $\sigma^2$ is the channel noise variance:
\begin{align*}
V^{(\infty)}_1 \stackrel{\Delta}{=} \lim_{t \rightarrow \infty} V^{(t)}_1 = \sigma^2(1-\alpha).
\end{align*}
\item The other variances approach zero:
\begin{align*}
V^{(\infty)}_i \stackrel{\Delta}{=} \lim_{t \rightarrow \infty} V^{(t)}_i = 0 
\end{align*}
for $i=2,3..d$.
\item The asymptotic convergence rate of all variances is exponential:
\begin{align*}
0 < \lim_{t \rightarrow \infty} \left|\frac{V^{(t)}_i - V^{(\infty)}_i}{\alpha^{t}}\right| < \infty 
\end{align*}
for $i=1,2..d$.
\item The zero approaching variances are upper bounded by the decaying exponential $\sigma^2\alpha^t$: 
\begin{align*}
V^{(t)}_i \leq \sigma^2\alpha^t 
\end{align*}
for $i=2,3..d$ and $t \geq 0$.
\end{enumerate}
%\item If $\alpha \geq 1$, then for $i=1,2...,d$ we have
%$\lim_{t \rightarrow \infty} V^{(t)}_i = 0$,
%such that
%$0 < \lim_{t \rightarrow \infty} V^{(t)}_i \cdot t < \infty$.
%\end{enumerate}
\end{theorem}
\begin{proof}
See Appendix \ref{var_app}.
\end{proof}

If $\alpha \geq 1$, the variances may still converge, but convergence rate may be as slow as $o(1/t)$, as illustrated in Appendix \ref{var_app}.

Convergence of the variances to zero implies that the Gaussians approach impulses. This is a desired property of the decoder, since the exact PDF that we want to calculate is indeed a weighted sum of impulses (see (\ref{exact_solution})).
It can be seen that by designing a code with $\alpha < 1$, i.e. $h_1^2 > \sum_{i=2}^d h_i^2$,
one variance
approaches a constant (and not zero). However, all the other
variances approach zero, where all variances converge in an exponential rate. This will be the
preferred mode because the information can be recovered even if a
single variance does not decay to zero, where exponential
convergence is certainly preferred over slow $1/t$
convergence. Therefore, from now on we shall restrict our analysis to magic square LDLC with $\alpha < 1$.

Theorem \ref{var_behavior} shows that every iteration, each variable node will generate $d-1$ messages with variances that approach zero, and a single message with variance that approaches a constant. The message with nonzero variance will be transmitted along the edge with largest weight (i.e. $h_1$). However, from the derivation of Appendix \ref{var_app} it can be seen that the opposite happens for the check nodes: each check node will generate $d-1$ messages with variances that approach a constant, and a single message with variance that approaches zero. The check node message with zero approaching variance will be transmitted along the edge with largest weight.

\subsection{Convergence of the Mean Values} \label{mean_convergence}
The reason that the messages are mixtures and not single Gaussians lies in the periodic extension step (\ref{check_message}) at the check nodes, and every Gaussian at the output of this step can be related to a single index of the infinite sum.
Therefore, we can label each Gaussian at iteration $t$ with a list of all the indices that were used in (\ref{check_message}) during its creation process in iterations $1,2,...t$. %This label should reflect the history of this Gaussian, and specifically the indices that were used to generate it in each check node and every iteration.

\begin{definition}[label of a Gaussian] \label{g_label}
The label of a Gaussian consists of a sequence of triplets of the form $\{t, c, i\}$, where $t$ is an iteration index, $c$ is a check node index and $i$ is an integer. The labels are initialized to the empty sequence. Then, the labels are updated along each iteration according to the following update rules:
\begin{enumerate}
\item In the periodic extension step (\ref{check_message}), each Gaussian in the output periodic mixture is assigned the label of the specific Gaussian of $p_j(x)$ that generated it, concatenated with a single triplet $\{t, c, i\}$, where $t$ is the current iteration index, $c$ is the check node index and $i$ is the index in the infinite sum of (\ref{check_message}) that corresponds to this Gaussian.
\item In the convolution step and the product step, each Gaussian in the output mixture is assigned a label that equals the concatenation of all the labels of the specific Gaussians in the input messages that formed this Gaussian.
%\item In the product step, each Gaussian in the product output mixture is assigned a label that equals the concatanation of all the labels of the specific Gaussians in the input messages that formed this Gaussian.
\item The stretching and normalization steps do not alter the label of each Gaussian: Each Gaussian in the stretched/normalized mixture inherits the label of the appropriate Gaussian in the original mixture.
\end{enumerate}
\end{definition}

\begin{definition}[a consistent Gaussian]
A Gaussian in a mixture is called ``[$t_a$, $t_b$] consistent'' if its label contains no contradictions for iterations $t_a$ to $t_b$, i.e. for every pair of triplets  $\{t_1, c_1, i_1\}$, $\{t_2, c_2, i_2\}$ such that $t_a \leq t_1, t_2 \leq t_b$, if $c_1=c_2$ then $i_1=i_2$. A [$0$, $\infty$] consistent Gaussian will be simply called a consistent Gaussian.
\end{definition}

We can relate every consistent Gaussian to a unique integer vector $\underline{b} \in \mathbb{Z}^n$, which holds the $n$ integers used in the $n$ check nodes. Since in the periodic extension step (\ref{check_message}) the sum is taken over all integers, a consistent Gaussian exists in each variable node message for every possible integer valued vector $\underline{b} \in \mathbb{Z}^n$. We shall see later that this consistent Gaussian corresponds to the lattice point $\boldsymbol{G}\underline{b}$.

According to Theorem \ref{var_behavior}, if we choose the nonzero values of $\boldsymbol{H}$ such that $\alpha < 1$, every variable node generates $d-1$ messages with variances approaching zero and a single message with variance that approaches a constant. We shall refer to these messages as ``narrow'' messages and ``wide'' messages, respectively. For a given integer valued vector $\underline{b}$, we shall concentrate on the consistent Gaussians that relate to $\underline{b}$ in all the $nd$ variable node messages that are generated in each iteration (a single Gaussian in each message).
The following lemmas summarize the asymptotic behavior of the mean values of these consistent Gaussians for the narrow messages.

\begin{lemma} \label{eq_mean_lemma}
For a magic square LDLC with degree $d$ and $\alpha < 1$, consider the $d-1$ narrow messages that are sent from a specific variable node. Consider further a single Gaussian in each message, which is the consistent Gaussian that relates to a given integer vector $\underline{b}$. Asymptotically, the mean values of these $d-1$ Gaussians become equal.
%Asymptotically, the $d-1$ mean values of the narrow consistent Gaussians that are sent from the same variable node relate to a given integer vector $\underline{b}$ become equal for all the $d-1$ narrow variable node messages that are .

\begin{proof}
See Appendix \ref{mean_app}.
\end{proof}
\end{lemma}

\begin{lemma} \label{narrow_mean_lemma}
For a magic square LDLC with dimension $n$, degree $d$ and $\alpha < 1$, consider only consistent Gaussians that relate to a given integer vector $\underline{b}$ and belong to narrow messages. Denote the common mean value of the $d-1$ such Gaussians that are sent from variable node $i$ at iteration $t$ by $m^{(t)}_i$, and arrange all these mean values in a column vector $\underline{m}^{(t)}$ of dimension $n$. Define the error vector $\underline{e}^{(t)} \stackrel{\Delta}{=} \underline{m}^{(t)} - \underline{x}$, where $\underline{x} = \boldsymbol{G} \underline{b}$ is the lattice point that corresponds to $\underline{b}$. Then, for large $t$, $\underline{e}^{(t)}$ satisfies:
\begin{eqnarray}  \label{thin_mean_recur}
\underline{e}^{(t+1)} \approx -\tilde{\boldsymbol{H}} \cdot \underline{e}^{(t)}
\end{eqnarray}
where $\tilde{\boldsymbol{H}}$ is derived from $\boldsymbol{H}$ by permuting the rows such that the $\pm h_1$ elements will be placed on the diagonal, dividing each row by the appropriate diagonal element ($h_1$ or $-h_1$), and then nullifying the diagonal.
\end{lemma}
\begin{proof}
See Appendix \ref{mean_app}.
\end{proof}

We can now state the following theorem, which describes the conditions for convergence and the steady state value of the mean values of the consistent Gaussians of the narrow variable node messages.
\begin{theorem} \label{narrow_mean_theorem}
For a magic square LDLC with $\alpha < 1$, the mean values of the consistent Gaussians of the narrow variable node messages that relate to a given integer vector $\underline{b}$ are assured to converge if and only if all the eigenvalues of $\tilde{\boldsymbol{H}}$ have magnitude less than $1$, where $\tilde{\boldsymbol{H}}$ is defined in Lemma \ref{narrow_mean_lemma}. When this condition is fulfilled, the mean values converge to the coordinates of the appropriate lattice point: $\underline{m}^{(\infty)} = \boldsymbol{G} \cdot \underline{b}$.
\end{theorem}
\begin{proof}
Immediate from Lemma \ref{narrow_mean_lemma}.
\end{proof}

%It can be seen that the mean values of the consistent Gaussians converge to the coordinates of the appropriate lattice point. However, as $t \rightarrow \infty$, the narrow variable node messages approach impulses, since their variance approaches zero. Also, we considered consistent Gaussians which correspond to a specific integer vector $\underline{b}$, but such a set of Gaussians exists for every possible choice of $\underline{b}$. Therefore, the narrow messages will have an impulse for every lattice point, located at the appropriate coordinate of the lattice point, which resembles (\ref{exact_solution}).

Note that without adding random signs to the LDLC nonzero values, the all-ones vector will be an eigenvector of $\tilde{\boldsymbol{H}}$ with eigenvalue $\frac{\sum_{i=2}^d h_i}{h_1}$, which may exceed $1$.

Interestingly, recursion (\ref{thin_mean_recur}) is also obeyed by
the error of the Jacobi method for solving systems of sparse
linear equations \cite{sparse} (see also Section \ref{enc}), when
it is used to solve $\boldsymbol{H}\underline{m}=\underline{b}$
(with solution $\underline{m}=\boldsymbol{G}\underline{b}$).
Therefore, the LDLC decoder can be viewed as a superposition of
Jacobi solvers, one for each possible value of the integer valued
vector $\underline{b}$.

We shall now turn to the convergence of the mean values of the wide messages. The asymptotic behavior is summarized in the following lemma.
\begin{lemma} \label{wide_mean_lemma}
For a magic square LDLC with dimension $n$ and $\alpha < 1$, consider only consistent Gaussians that relate to a given integer vector $\underline{b}$ and belong to wide messages. Denote the mean value of such a Gaussian that is sent from variable node $i$ at iteration $t$ by $m^{(t)}_i$, and arrange all these mean values in a column vector $\underline{m}^{(t)}$ of dimension $n$. Define the error vector $\underline{e}^{(t)} \stackrel{\Delta}{=} \underline{m}^{(t)} - \boldsymbol{G}\underline{b}$. Then, for large $t$, $\underline{e}^{(t)}$ satisfies:
%Denote the mean value of the wide message of the $i$'th variable node at iteration $t$ by $m^{(t)}_i$, and the appropriate error by $e^{(t)}_i \stackrel{\Delta}{=} m^{(t)}_i - x_i$, where $\underline{x}=G\underline{b}$. Arrange all the error values in a column vector $\underline{e}^{(t)}$ of dimension $n$. Then, for large $t$, $\underline{e}^{(t)}$ satisfies:
\begin{eqnarray} \label{wide_mean_recur}
\underline{e}^{(t+1)} \approx -\boldsymbol{F} \cdot \underline{e}^{(t)} + (1-\alpha) (\underline{y}-\boldsymbol{G}\underline{b})
\end{eqnarray}
where $\underline{y}$ is the noisy codeword and $\boldsymbol{F}$ is an $n \times n$ matrix defined by:
\begin{align}
F_{k,l} = \left\{ \begin{array}{ll}
\frac{H_{r,k}}{H_{r,l}} & \textrm{if } k \neq l \textrm{ and there exist a row } r \textrm{ of H} \\
& \textrm{for which } |H_{r,l}|=h_1 \textrm{ and } H_{r,k} \neq 0 \\
0 & \textrm{otherwise}
\end{array} \right.
\end{align}

%\begin{align}
%F_{k,l} = \left\{ \begin{array}{ll}
%\frac{H_{r,k}}{H_{r,l}} & \textrm{if exist a row } r \textrm{ of H for which } \\
%& |H_{r,l}|=h_1 \textrm{ and } H_{r,k} \neq 0 \\
%0 & otherwise
%\end{array} \right.
%\end{align}

%The matrix $F$ can be constructed from $\boldsymbol{H}$ as follows. To construct the $k$'th row of $F$, denote by $r_i$, $i=1,2,...d$, the index of the element in the $k$'th column of $\boldsymbol{H}$ with value $h_i$ (i.e. $|H_{r_i, k}| = h_i$). Denote by $l_i$, $i=1,2,...d$, the index of the element in the $r_i$'th row of $\boldsymbol{H}$ with value $h_1$ (i.e. $|H_{r_i, l_i}| = h_1$). The $k$'th row of $F$ will then be all zeros, except for the $d-1$ locations $l_i$, $i=2,3...d$, where $F_{k, l_i} = \frac{H_{r_i, k}}{H_{r_i, l_i}}$.
\end{lemma}
\begin{proof}
See Appendix \ref{mean_app}, where an alternative way to construct $\boldsymbol{F}$ from $\boldsymbol{H}$ is also presented.
\end{proof}

The conditions for convergence and steady state solution for the wide messages are described in the following theorem.
\begin{theorem} \label{wide_mean_theorem}
%The recursion (\ref{wide_mean_recur}) is assured to converge
For a magic square LDLC with $\alpha < 1$, the mean values of the consistent Gaussians of the wide variable node messages that relate to a given integer vector $\underline{b}$ are assured to converge if and only if all the eigenvalues of $\boldsymbol{F}$ have magnitude less than $1$, where $\boldsymbol{F}$ is defined in Lemma \ref{wide_mean_lemma}. When this condition is fulfilled, the steady state solution is $\underline{m}^{(\infty)} = \boldsymbol{G} \cdot \underline{b} + (1-\alpha)(\boldsymbol{I}+\boldsymbol{F})^{-1} (\underline{y}-\boldsymbol{G} \cdot \underline{b})$.
\end{theorem}
\begin{proof}\
Immediate from Lemma \ref{wide_mean_lemma}.
\end{proof}

Unlike the narrow messages, the mean values of the wide messages do not converge to the appropriate lattice point coordinates. The steady state error depends on the difference between the noisy observation and the lattice point, as well as on $\alpha$, and it decreases to zero as $\alpha \rightarrow 1$. Note that the final PDF of a variable is generated by multiplying \emph{all} the $d$ check node messages that arrive to the appropriate variable node. $d-1$ of these messages are wide, and therefore their mean values have a steady state error. One message is narrow, so it converges to an impulse at the lattice point coordinate. Therefore, the final product will be an impulse at the correct location, where the wide messages will only affect the magnitude of this impulse. As long as the mean values errors are not too large (relative to the width of the wide messages), this should not cause an impulse that corresponds to a wrong lattice point to have larger amplitude than the correct one. However, for large noise, these steady state errors may cause the decoder to deviate from the ML solution (As explained in Section \ref{amp_convergence}).

To summarize the results for the mean values, we considered the mean values of all the consistent Gaussians that correspond to a given integer vector $\underline{b}$. A single Gaussian of this form exists in each of the $nd$ variable node messages that are generated in each iteration. For each variable node, $d-1$ messages are narrow (have variance that approaches zero) and a single message is wide (variance approaches a constant). Under certain conditions on $\boldsymbol{H}$, the mean values of all the narrow messages converge to the appropriate coordinate of the lattice point $\boldsymbol{G}\underline{b}$. Under additional conditions on $\boldsymbol{H}$, the mean values of the wide messages converge, but the steady state values contain an error term.

We analyzed the behavior of consistent Gaussian. It should be
noted that there are many more non-consistent Gaussians.
Furthermore non-consistent Gaussians are generated in each
iteration for any existing consistent Gaussian. We conjecture that
unless a Gaussian is consistent, or becomes consistent along the
iterations, it fades out, at least at noise conditions where the
algorithm converges. The reason is that non-consistency in the
integer values leads to mismatch in the corresponding PDF's, and
so the amplitude of that Gaussian is attenuated.

We considered consistent Gaussians which correspond to a specific integer vector $\underline{b}$, but such a set of Gaussians exists for every possible choice of $\underline{b}$, i.e. for every lattice point. Therefore, the narrow messages will converge to  a solution that has an impulse at the appropriate coordinate of every lattice point. This resembles the exact solution (\ref{exact_solution}), so the key for proper convergence lies in the amplitudes: we would like the consistent Gaussians of the ML lattice point to have the largest amplitude for each message. %(note that the recursion for the narrow messages will converge for any initial conditions. Therefore, Gaussians that became consistent only from some iteration $t_0$ will also converge to the appropriate lattice point. The Gaussian at each lattice point will then be the sum of all the Gaussians that started being consistent at some iteration).
%However, the analysis of the amplitudes is more complex and was not yet finalized.
%(note that we should also show that the noise term due to non-consistent Gaussians does not cause other Gaussians to exceed the amplitude of the ML Gaussians).

\subsection{Convergence of the Amplitudes} \label{amp_convergence}
We shall now analyze the behavior of the amplitudes of consistent Gaussians (as discussed later, this is not enough for complete convergence analysis, but it certainly gives insight to the nature of the convergence process and its properties). The behavior of the amplitudes of consistent Gaussians is summarized in the following lemma.
\begin{lemma} \label{amp_lemma}
For a magic square LDLC with dimension $n$, degree $d$ and $\alpha < 1$, consider the $nd$ consistent Gaussians that relate to a given integer vector $\underline{b}$ in the variable node messages that are sent at iteration $t$ (one consistent Gaussian per message). Denote the amplitudes of these Gaussians by $p^{(t)}_i$, $i=1,2,...nd$, and define the log-amplitude as $l^{(t)}_i = \log p^{(t)}_i$. Arrange these $nd$ log-amplitudes in a column vector $\underline{l}^{(t)}$, such that element $(k-1)d+i$ corresponds to the message that is sent from variable node $k$ along an edge with weight $\pm h_i$. Assume further that the bipartite graph of the LDLC contains no 4-loops. Then, the log-amplitudes satisfy the following recursion:
\begin{align} \label{amp_recur}
\underline{l}^{(t+1)} = \boldsymbol{A} \cdot \underline{l}^{(t)} - \underline{a}^{(t)} - \underline{c}^{(t)}
\end{align}
with initialization $\underline{l}^{(0)} = \underline{0}$. $\boldsymbol{A}$ is an $nd \times nd$ matrix which is all zeros except exactly $(d-1)^2$ '$1$'s in each row and each column.
The element of the excitation vector $\underline{a}^{(t)}$ at location $(k-1)d+i$ (where $k=1,2,...n$ and $i=1,2,...d$) equals:
\begin{align} \label{excitation}
a^{(t)}_{(k-1)d+i} =
\end{align}
\begin{align*}
=\frac{\hat{V}^{(t)}_{k,i}}{2} \left(\sum_{\substack{l=1 \\ l \neq i}}^{d} \sum_{\substack{j=l+1 \\ j \neq i}}^{d} \frac{\left(\tilde{m}^{(t)}_{k,l}-\tilde{m}^{(t)}_{k,j}\right)^2}{\tilde{V}^{(t)}_{k,l} \cdot \tilde{V}^{(t)}_{k,j}} + \sum_{\substack{l=1 \\ l \neq i}}^{d} \frac{\left(\tilde{m}^{(t)}_{k,l}-y_k\right)^2}{\sigma^2 \cdot \tilde{V}^{(t)}_{k,l}}\right)
\end{align*}
%The $i$'th element of the excitation vector $\underline{a}^{(t)}$ (where $i=1,2,...nd$) equals:
%\begin{align} \label{excitation}
%a^{(t)}_i = \frac{\hat{V}}{2} \left(\sum_{l=1}^{d-1} \sum_{j=l+1}^{d-1} \frac{(\tilde{m}_l-\tilde{m}_j)^2}{\tilde{V}_l \cdot \tilde{V}_j} + \sum_{l=1}^{d-1}\frac{(\tilde{m}_l-y_k)^2}{\sigma^2 \cdot \tilde{V}_l}\right)
%\end{align}
%where $\tilde{m}_1, \tilde{m}_2,...,\tilde{m}_{d-1}$ denote the $d-1$ mean values of the appropriate consistent Gaussians of the check node messages that are used to generate the $i$'th variable node message at iteration $t$ (for convenience of notations, the $i$ and $t$ indices were omitted). $\tilde{V}_1, \tilde{V}_2,...,\tilde{V}_{d-1}$ denote the variances of these messages, respectively, and $\hat{V} \stackrel{\Delta}{=} \left(\frac{1}{\sigma^2} + \sum_{i=1}^{d-1}\frac{1}{\tilde{V}_i}\right)^{-1}$. $y_k$ is the noisy channel observation that relates to the variable node that generates the $i$'th message. Finally, $\underline{k}^{(t)}$ is a constant excitation term that is independent of the integer vector $\underline{b}$ (i.e. is the same for all consistent Gaussians).
where $\tilde{m}^{(t)}_{k,l}$ and $\tilde{V}^{(t)}_{k,l}$ denote the mean value and variance of the consistent Gaussian that relates to the integer vector $\underline{b}$ in the check node message that arrives to variable node $k$ at iteration $t$ along an edge with weight $\pm h_l$. $y_k$ is the noisy channel observation of variable node $k$, and $\hat{V}^{(t)}_{k,i} \stackrel{\Delta}{=} \left(\frac{1}{\sigma^2} + \sum_{\substack{l=1 \\ l \neq i}}^{d}\frac{1}{\tilde{V}^{(t)}_{k,l}}\right)^{-1}$. Finally, $\underline{c}^{(t)}$ is a constant excitation term that is independent of the integer vector $\underline{b}$ (i.e. is the same for all consistent Gaussians). Note that an iteration is defined as sending variable node messages, followed by sending check node messages. The first iteration (where the variable nodes send the initialization PDF) is regarded as iteration $0$.
\end{lemma}
\begin{proof}
At the check node, the amplitude of a Gaussian at the convolution output is the product of the amplitudes of the corresponding Gaussians in the appropriate variable node messages. At the variable node, the amplitude of a Gaussian at the product output is the product of the amplitudes of the corresponding Gaussians in the appropriate check node messages, multiplied by the Gaussian scaling term, according to claim \ref{product_claim}. Since we assume that the bipartite graph of the LDLC contains no 4-loops, an amplitude of a variable node message at iteration $t$ will therefore equal the product of $(d-1)^2$ amplitudes of Gaussians of variable node messages from iteration $t-1$, multiplied by the Gaussian scaling term. This proves (\ref{amp_recur}) and shows that $A$ has $(d-1)^2$ '1's in every row. However, since each variable node message affects exactly $(d-1)^2$ variable node messages of the next iteration, $A$ must also have $(d-1)^2$ '1's in every column. The total excitation term $-\underline{a}^{(t)} - \underline{c}^{(t)}$ corresponds to the logarithm of the Gaussian scaling term.
Each element of this scaling term results from the product of $d-1$ check node Gaussians and the channel Gaussian, according to claim \ref{product_claim}. This scaling term sums over all the pairs of Gaussians, and in (\ref{excitation}) the sum is separated to pairs that include the channel Gaussian and pairs that do not. The total excitation is divided between (\ref{excitation}), which depends on the choice of the integer vector $\underline{b}$, and $\underline{c}^{(t)}$, which includes all the constant terms that are independent on $\underline{b}$ (including the normalization operation which is performed at the variable node).
\end{proof}
  %In the coming analysis we shall ignore the normalization step (which forces the sum of all the amplitudes of Gaussians of the same message to be 1), since it does not change the relative amplitude between two Gaussians of the same message (and in principle can be applied only at the end of the algorithm and not every iteration).

Since there are exactly $(d-1)^2$ '$1$'s in each column of the matrix $\boldsymbol{A}$, it is easy to see that the all-ones vector is an eigenvector of $\boldsymbol{A}$, with eigenvalue $(d-1)^2$. If $d>2$, this eigenvalue is larger than $1$, meaning that the recursion (\ref{amp_recur}) is non-stable.

It can be seen that the excitation term $\underline{a}^{(t)}$ has two components. The first term sums the squared differences between the mean values of all the possible pairs of received check node messages (weighted by the inverse product of the appropriate variances). It therefore measures the mismatch between the incoming messages. This mismatch will be small if the mean values of the consistent Gaussians converge to the coordinates of a lattice point (\emph{any} lattice point). The second term sums the squared differences between the mean values of the incoming messages and the noisy channel output $y_k$. This term measures the mismatch between the incoming messages and the channel measurement. It will be smallest if the mean values of the consistent Gaussians converge to the coordinates of the ML lattice point.

The following lemma summarizes some properties of the excitation term $\underline{a}^{(t)}$.
\begin{lemma} \label{excitation_behavior}
For a magic square LDLC with dimension $n$, degree $d$, $\alpha<1$ and no 4-loops, consider the consistent Gaussians that correspond to a given integer vector $\underline{b}$. According to Lemma \ref{amp_lemma}, their amplitudes satisfy recursion (\ref{amp_recur}). The excitation term $\underline{a}^{(t)}$ of (\ref{amp_recur}), which is defined by (\ref{excitation}), satisfies the following properties:
\begin{enumerate}
\item $a^{(t)}_i$, the $i$'th element of $\underline{a}^{(t)}$, is non-negative, finite and bounded for every $i$ and every $t$. %, i.e. there exists a constant $a_m$ such that $|a^{(t)}_i| \leq a_m$ for all $1 \leq i\ leq nd$ and $t >0$.
Moreover, $a^{(t)}_i$ converges to a finite non-negative steady state value as $t \rightarrow \infty$.
\item $\lim_{t \rightarrow \infty} \sum_{i=1}^{nd} a^{(t)}_i = \frac{1}{2\sigma^2}(\boldsymbol{G} \underline{b} - \underline{y})^T \boldsymbol{W} (\boldsymbol{G} \underline{b} - \underline{y})$,
where $\underline{y}$ is the noisy received codeword and $\boldsymbol{W}$ is a positive definite matrix defined by:
\begin{align} \label{w_def}
\boldsymbol{W} \stackrel{\Delta}{=}
(d+1-\alpha)\boldsymbol{I} %- (1-\alpha)(I+F^{-1})^{{-1}^T}(I+F^{-1})^{-1} +
-2(1-\alpha) (\boldsymbol{I}+\boldsymbol{F})^{-1} +
\end{align}
\begin{align*}
+(1-\alpha)(\boldsymbol{I}+\boldsymbol{F})^{{-1}^T}\left((d-1)^2 \boldsymbol{I} -\boldsymbol{F}^T \boldsymbol{F}\right)(\boldsymbol{I}+\boldsymbol{F})^{-1}
\end{align*}
where $\boldsymbol{F}$ is defined in Lemma \ref{wide_mean_lemma}.
\item For an LDLC with degree $d>2$, the weighted infinite sum $\sum_{j=0}^{\infty} \frac{\sum_{i=1}^{nd} a^{(j)}_i}{(d-1)^{2j+2}}$ converges to a finite value.
\end{enumerate}
\end{lemma}
\begin{proof}
See Appendix \ref{amp_rec_app}.
\end{proof}

The following theorem addresses the question of which consistent Gaussian will have the maximal asymptotic amplitude. We shall first consider the case of an LDLC with degree $d>2$, and then consider the special case of $d=2$ in a separate theorem.
\begin{theorem} \label{d_gt_2_amp_theorem}
For a magic square LDLC with dimension $n$, degree $d>2$, $\alpha<1$ and no 4-loops, consider the $nd$ consistent Gaussians that relate to a given integer vector $\underline{b}$ in the variable node messages that are sent at iteration $t$ (one consistent Gaussian per message). Denote the amplitudes of these Gaussians by $p^{(t)}_i$, $i=1,2,...nd$, and define the product-of-amplitudes as $P^{(t)}\stackrel{\Delta}{=}\prod_{i=1}^{nd} p^{(t)}_i$.
Define further $S = \sum_{j=0}^{\infty} \frac{\sum_{i=1}^{nd} a^{(j)}_i}{(d-1)^{2j+2}}$, where $a^{(j)}_i$ is defined by (\ref{excitation}) ($S$ is well defined according to Lemma \ref{excitation_behavior}). Then:
\begin{enumerate}
%\item The infinite sum that defines $S$ converges for every finite integer vector $\underline{b}$.
\item The integer vector $\underline{b}$ for which the consistent Gaussians will have the largest asymptotic product-of-amplitudes $\lim_{t \rightarrow \infty} P^{(t)}$ is the one for which $S$ is minimized.
\item The product-of-amplitudes for the consistent Gaussians that correspond to all other integer vectors will decay to zero in a super-exponential rate.
\end{enumerate}
\end{theorem}

\begin{proof}
As in Lemma \ref{amp_lemma}, define the log-amplitudes $l^{(t)}_i \stackrel{\Delta}{=} \log p^{(t)}_i$. Define further $s^{(t)}\stackrel{\Delta}{=}  \sum_{i=1}^{nd} l^{(t)}_i$. Taking the element-wise sum of (\ref{amp_recur}), we get:
\begin{align} \label{sum_amp_recur}
s^{(t+1)} = (d-1)^2 s^{(t)} - \sum_{i=1}^{nd} a^{(t)}_i
\end{align}
with initialization $s^{(0)}=0$. Note that we ignored the term $\sum_{i=1}^{nd}c^{(t)}_i$. As shown below, we are looking for the vector $\underline{b}$ that maximizes $s^{(t)}$. Since (\ref{sum_amp_recur}) is a linear difference equation, and the term $\sum_{i=1}^{nd}c^{(t)}_i$ is independent of $\underline{b}$, its effect on $s^{(t)}$ is common to all $\underline{b}$ and is therefore not interesting.

Define now $\tilde{s}^{(t)}\stackrel{\Delta}{=} \frac{s^{(t)}}{(d-1)^{2t}} $. Substituting in (\ref{sum_amp_recur}), we get:
\begin{align} \label{norm_sum_amp_recur}
\tilde{s}^{(t+1)} =  \tilde{s}^{(t)} - \frac{1}{(d-1)^{2t+2}}\sum_{i=1}^{nd} a^{(t)}_i
\end{align}
with initialization $\tilde{s}^{(0)}=0$, which can be solved to get:
\begin{align} \label{norm_sum_amp_solution}
\tilde{s}^{(t)} = - \sum_{j=0}^{t-1} \frac{\sum_{i=1}^{nd} a^{(j)}_i}{(d-1)^{2j+2}}
\end{align}
We would now like to compare the amplitudes of consistent Gaussians with various values of the corresponding integer vector $\underline{b}$ in order to find the lattice point whose consistent Gaussians will have largest product-of-amplitudes.
From the definitions of $s^{(t)}$ and $\tilde{s}^{(t)}$ we then have:
\begin{eqnarray}
P^{(t)}=e^{s^{(t)}}=e^{(d-1)^{2t} \cdot \tilde{s}^{(t)}}
\end{eqnarray}
Consider two integer vectors $\underline{b}$ that relate to two lattice points. Denote the corresponding product-of-amplitudes by $P_0^{(t)}$ and $P_1^{(t)}$, respectively, and assume that for these two vectors $S$ converges to the values $S_0$ and $S_1$, respectively.
Then, taking into account that $\lim_{t \rightarrow \infty} \tilde{s}^{(t)} = -S$, the asymptotic ratio of the product-of-amplitudes for these lattice points will be:
\begin{eqnarray} \label{ratio_d_ge_2}
\lim_{t \rightarrow \infty} \frac{P_1^{(t)}}{P_0^{(t)}}=\frac{e^{-(d-1)^{2t} \cdot S_1}}{e^{-(d-1)^{2t} \cdot S_0}}=e^{(d-1)^{2t} \cdot (S_0-S_1)}
\end{eqnarray}
It can be seen that if $S_0 < S_1$, the ratio decreases to zero in a super exponential rate. This shows that the lattice point for which $S$ is minimized will have the largest product-of-amplitudes, where for all other lattice points, the product-of-amplitudes will decay to zero in a super-exponential rate (recall that the normalization operation at the variable node keeps the sum of all amplitudes in a message to be 1). %The $k^{(t)}_i$ of (\ref{excitation})can be ignored in the definition of $S$ since it results in a term which is common to all consistent Gaussians (independent of $\underline{b}$), so it does not change the relative amplitudes of two lattice points.
This completes the proof of the theorem.
\end{proof}

We now have to find which integer valued vector $\underline{b}$ minimizes $S$. The analysis is difficult because the weighting factor inside the sum of (\ref{norm_sum_amp_solution}) performs exponential weighting of the excitation terms, where the dominant terms are those of the first iterations. Therefore, we can not use the asymptotic results of Lemma \ref{excitation_behavior}, but have to analyze the transient behavior. However, the analysis is simpler for the case of an LDLC with row and column degree of $d=2$, so we shall first turn to this simple case (note that for this case, both the convolution in the check nodes and the product at the variable nodes involve only a single message).

\begin{theorem} \label{d_eq_2_amp_theorem}
For a magic square LDLC with dimension $n$, degree $d=2$, $\alpha<1$ and no 4-loops, consider the $2n$ consistent Gaussians that relate to a given integer vector $\underline{b}$ in the variable node messages that are sent at iteration $t$ (one consistent Gaussian per message). Denote the amplitudes of these Gaussians by $p^{(t)}_i$, $i=1,2,...2n$, and define the product-of-amplitudes as $P^{(t)}\stackrel{\Delta}{=}\prod_{i=1}^{2n} p^{(t)}_i$.
Then:
\begin{enumerate}
\item The integer vector $\underline{b}$ for which the consistent Gaussians will have the largest asymptotic product-of-amplitudes $\lim_{t \rightarrow \infty} P^{(t)}$ is the one for which $(\boldsymbol{G} \underline{b} - \underline{y})^T \boldsymbol{W} (\boldsymbol{G} \underline{b} - \underline{y})$ is minimized, where $\boldsymbol{W}$ is defined by (\ref{w_def}) and $\underline{y}$ is the noisy received codeword.
\item The product-of-amplitudes for the consistent Gaussians that correspond to all other integer vectors will decay to zero in an exponential rate.
\end{enumerate}
\end{theorem}
\begin{proof}
For $d=2$ (\ref{sum_amp_recur}) becomes:
\begin{align} \label{sum_amp_recur_2}
s^{(t+1)} = s^{(t)} - \sum_{i=1}^{2n} a^{(t)}_i
\end{align}
With solution:
\begin{align} \label{norm_sum_amp_solution_d_2}
s^{(t)} = - \sum_{j=0}^{t-1} \sum_{i=1}^{2n} a^{(j)}_i
\end{align}
%It can be seen that $s^{(t)}$ is the negated accumulated sum of the excitation sum term $\sum_{i=1}^{nd} a^{(j)}_i$ over iterations $j=1$ to $t$. As shown in the proof of theorem \ref{d_gt_2_amp_theorem}, the excitation term elements approach a steady state value, so
Denote $S_a = \lim _{j \rightarrow \infty} \sum_{i=1}^{2n} a^{(j)}_i$. $S_a$ is well defined according to Lemma \ref{excitation_behavior}. For large $t$, we then have $s^{(t)} \approx -t \cdot S_a$. Therefore, for two lattice points with excitation sum terms which approach $S_{a0}, S_{a1}$, respectively, the ratio of the corresponding product-of-amplitudes will approach
\begin{eqnarray} \label{ratio_d_eq_2}
\lim_{t \rightarrow \infty} \frac{P_1^{(t)}}{P_0^{(t)}} = \frac{e^{-S_{a1} \cdot t}}{e^{-S_{a0} \cdot t}}=e^{ (S_{a0}-S_{a1}) \cdot t}
\end{eqnarray}
If $S_{a0} < S_{a1}$, the ratio decreases to zero exponentially (unlike the case of $d>2$ where the rate was super-exponential, as in (\ref{ratio_d_ge_2})). This shows that the lattice point for which $S_a$ is minimized will have the largest product-of-amplitudes, where for all other lattice points, the product-of-amplitudes will decay to zero in an exponential rate (recall that the normalization operation at the variable node keeps the sum of all amplitudes in a message to be 1). This completes the proof of the second part of the theorem.

We still have to find the vector $\underline{b}$ that minimizes $S_a$. The basic difference between the case of $d=2$ and the case of $d>2$ is that for $d>2$ we need to analyze the transient behavior of the excitation terms, where for $d=2$ we only need to analyze the asymptotic behavior, which is much easier to handle.

%In appendix \ref{amp_rec_app}, the asymptotic behavior of the excitation terms is analyzed. It is shown that for a given integer vector $\underline{b}$, the corresponding asymptotic excitation sum term equals:
According to Lemma \ref{excitation_behavior}, we have:
\begin{align} \label{steady_state_excitation}
S_a \stackrel{\Delta}{=} \lim _{j \rightarrow \infty} \sum_{i=1}^{2n} a^{(j)}_i = \frac{1}{2\sigma^2}(\boldsymbol{G} \underline{b} - \underline{y})^T \boldsymbol{W} (\boldsymbol{G} \underline{b} - \underline{y})
\end{align}
where $\boldsymbol{W}$ is defined by (\ref{w_def}) and $\underline{y}$ is the noisy received codeword. Therefore, for $d=2$, the lattice points whose consistent Gaussians will have largest product-of-amplitudes is the point for which $(\boldsymbol{G} \underline{b} - \underline{y})^T \boldsymbol{W} (\boldsymbol{G} \underline{b} - \underline{y})$ is minimized. This completes the proof of the theorem.
\end{proof}

For $d=2$ we could find an explicit expression for the ``winning'' lattice point.
As discussed above, we could not find an explicit expression for $d>2$, since the result depends on the transient behavior of the excitation sum term, and not only on the steady state value. However, a reasonable conjecture is to assume that $\underline{b}$ that maximizes the steady state excitation will also maximize the term that depends on the transient behavior. %As shown in appendix \ref{amp_rec_app}, the steady state excitation for $d>2$ can also be expressed in the form of (\ref{steady_state_excitation}).
This means that a reasonable conjecture is to assume that the ``winning'' lattice point for $d>2$ will also minimize an expression of the form (\ref{steady_state_excitation}).

Note that for $d>2$ we can still show that for ``weak'' noise, the ML point will have the minimal $S$. To see that, it comes out from (\ref{excitation}) that for zero noise, the ML lattice point will have $a^{(t)}_i=0$ for every $t$ and $i$, where all other lattice points will have $a^{(t)}_i>0$ for at least some $i$ and $t$. Therefore, the ML point will have a minimal excitation term along the transient behavior so it will surely have the minimal $S$ and the best product-of-amplitudes. As the noise increases, it is difficult to analyze the transient behavior of $a^{(t)}_i$, as discussed above.

Note that the ML solution minimizes $(\boldsymbol{G} \underline{b} - \underline{y})^T (\boldsymbol{G} \underline{b} - \underline{y})$, where the above analysis yields minimization of $(\boldsymbol{G} \underline{b} - \underline{y})^T \boldsymbol{W} (\boldsymbol{G} \underline{b} - \underline{y})$. Obviously, for zero noise (i.e. $\underline{y}=\boldsymbol{G} \cdot\ \underline{b}$) both minimizations will give the correct solution with zero score. As the noise increases, the solutions may deviate from one another. Therefore, both minimizations will give the same solution for ``weak'' noise but may give different solutions for ``strong'' noise.

An example for another decoder that performs this form of minimization is the linear detector, which calculates $\hat{\underline{b}}=\left\lfloor \boldsymbol{H} \cdot \underline{y} \right\rceil$ (where $\left\lfloor x \right\rceil$ denotes the nearest integer to $x$). This is equivalent to minimizing $(\boldsymbol{G} \underline{b} - \underline{y})^T \boldsymbol{W} (\boldsymbol{G} \underline{b} - \underline{y})$ with $\boldsymbol{W}=\boldsymbol{H}^T \boldsymbol{H} = \boldsymbol{G}^{{-1}^T} \boldsymbol{G}^{-1}$. %$M=H=G^{-1}$.
The linear detector fails to yield the ML solution if the noise is too strong, due to its inherent noise amplification.

For the LDLC iterative decoder, we would like that the deviation from the ML decoder due to the $\boldsymbol{W}$ matrix would be negligible in the expected range of noise variance. Experimental results (see Section \ref{sim_res}) show that the iterative decoder indeed converges to the ML solution for noise variance values that approach channel capacity. However, for quantization or shaping applications (see Section \ref{shaping}), where the effective noise is uniformly distributed along the Voronoi cell of the lattice (and is much stronger than the noise variance at channel capacity) the iterative decoder fails, and this can be explained by the influence of the $\boldsymbol{W}$ matrix on the minimization, as described above. Note from (\ref{w_def}) that as $\alpha \rightarrow 1$, $\boldsymbol{W}$ approaches a scaled identity matrix, which means that the minimization criterion approaches the ML criterion. However, the variances converge as $\alpha^t$, so as $\alpha \rightarrow 1$ convergence time approaches infinity.

Until this point, we concentrated only on consistent Gaussians, and checked what lattice point maximizes the product-of-amplitudes of all the corresponding consistent Gaussians. However, this approach does not necessarily lead to the lattice point that will be finally chosen by the decoder, due to 3 main reasons:
\begin{enumerate}
\item It comes out experimentally that the strongest Gaussian in each message is not necessarily a consistent Gaussian, but a Gaussian that started as non-consistent and became consistent at a certain iteration. %the non-consistent Gaussians may have significant influence, especially at the first iterations, where they may be stronger than the consistent ones (recall that for $d>2$ the ``winning'' Gaussians depend mainly on the behavior at the first iterations). It is reasonable to assume that the non-consistent Gaussians will fade away if they keep being non-consistent; however,
Such a Gaussian %a non-consistent Gaussian that starts to be consistent at some iteration
will finally converge to the appropriate lattice point, since the convergence of the mean values is independent of initial conditions. The non-consistency at the first several iterations, where the mean values are still very noisy, allows these Gaussians to accumulate stronger amplitudes than the consistent Gaussians (recall that the exponential weighting in (\ref{norm_sum_amp_solution}) for $d>2$ results in strong dependency on the behavior at the first iterations).
\item There is an exponential number of Gaussians that start as non-consistent and become consistent (with the same integer vector $\underline{b}$) at a certain iteration, and the final amplitude of the Gaussians at the lattice point coordinates will be determined by the sum of all these Gaussians.  % that started being consistent at some iteration with the same integer vector $\underline{b}$ (where probably some of these will be stronger than the corresponding consistent Gaussian itself, due to better excitation terms in the first iterations).
\item We ignored non-consistent Gaussians that endlessly remain non-consistent. We have not shown it analytically, but it is reasonable to assume that the excitation terms for such Gaussians will be weaker than for Gaussians that become consistent at some point, so their amplitude will fade away to zero. However, non-consistent Gaussians are born every iteration, even at steady state. The ``newly-born'' non-consistent Gaussians may appear as sidelobes to the main impulse, since it may take several iterations until they are attenuated. Proper choice of the coefficients of $\boldsymbol{H}$ may minimize this effect, as discussed in Sections \ref{decoder} and \ref{choose_seq}. However, these Gaussians may be a problem for small $d$ (e.g. $d=2$) where the product step at the variable node does not include enough messages to suppress them.
%\item Finally, we only considered maximizing the product-of-amplitudes over all the messages, but this does not guarantee that the appropriate Gaussian will be the stronger in each message.
\end{enumerate}
Note that the first two issues are not a problem for $d=2$, where the winning lattice point depends only on the asymptotic behavior. The amplitude of a sum of Gaussians that converged to the same coordinates will still be governed by (\ref{norm_sum_amp_solution_d_2}) and the winning lattice point will still minimize (\ref{steady_state_excitation}). The third issue is a problem for small $d$, but less problematic for large $d$, as described above.

As a result, we can not regard the convergence analysis of the consistent Gaussians' amplitudes   as a complete convergence analysis. However, it can certainly be used as a qualitative analysis that gives certain insights to the convergence process. Two main observations are:
\begin{enumerate}
\item The narrow variable node messages tend to converge to single impulses at the coordinates of a single lattice point. This results from (\ref{ratio_d_ge_2}), (\ref{ratio_d_eq_2}), which show that the ``non-winning'' consistent Gaussians will have amplitudes that decrease to zero relative to the amplitude of the ``winning'' consistent Gaussian. This result remains valid for the sum of non-consistent Gaussians that became consistent at a certain point, %, and generate impulses at the same locations as those of the appropriate consistent Gaussian,
because it results from the non-stable nature of the recursion (\ref{amp_recur}), which makes strong Gaussians stronger in an exponential manner. The single impulse might be accompanied by weak ``sidelobes'' due to newly-born non-consistent Gaussians.

Interestingly, this form of solution is different from the exact solution (\ref{exact_solution}), where every lattice point is represented by an impulse at the appropriate coordinate, with amplitude that depends on the Euclidean distance of the lattice point from the observation. The iterative decoder's solution has a single impulse that corresponds to a single lattice point, where all other impulses have amplitudes that decay to zero. This should not be a problem, as long as the ML point is the remaining point (see discussion above).

\item  We have shown that for $d=2$ the strongest consistent Gaussians relate to $\underline{b}$ that minimizes an expression of the form $(\boldsymbol{G} \underline{b} - \underline{y})^T \boldsymbol{W} (\boldsymbol{G} \underline{b} - \underline{y})$. We proposed a conjecture that this is also true for $d>2$. We can further widen the conjecture to say that the finally decoded $\underline{b}$ (and not only the $\underline{b}$ that relates to strongest consistent Gaussians) minimizes such an expression. Such a conjecture can explain why the iterative decoder works well for decoding near channel capacity, but fails for quantization or shaping, where the effective noise variance is much larger.
\end{enumerate}

\subsection{Summary of Convergence Results} \label{summary_convergence}
To summarize the convergence analysis, it was first shown that the
variable node messages are Gaussian mixtures. Therefore, it is
sufficient to analyze the sequences of variances, mean values and
relative amplitudes of the Gaussians in each mixture. Starting
with the variances, it was shown that with proper choice of the
magic square LDLC generating sequence, each variable node
generates $d-1$ ``narrow'' messages, whose variance decreases
exponentially to zero, and a single ``wide'' message, whose
variance reaches a finite value. Consistent Gaussians were then
defined as Gaussians that their generation process always involved
the same integer at the same check node. Consistent Gaussians can
then be related to an integer vector $\underline{b}$ or
equivalently to the lattice point $\boldsymbol{G} \underline{b}$.
It was then shown that under appropriate conditions on
$\boldsymbol{H}$, the mean values of consistent Gaussians that
belong to narrow messages converge to the coordinates of the
appropriate lattice point. The mean values of wide messages also
converge to these coordinates, but with a steady state error.
Then, the amplitudes of consistent Gaussians were analyzed. For
$d=2$ it was shown that the consistent Gaussians with maximal
product-of-amplitudes (over all messages) are those that
correspond to an integer vector $\underline{b}$ than minimizes
$(\boldsymbol{G} \underline{b} - \underline{y})^T \boldsymbol{W}
(\boldsymbol{G} \underline{b} - \underline{y})$, where
$\boldsymbol{W}$ is a positive definite matrix that depends only
on $\boldsymbol{H}$. The product-of-amplitudes for all other
consistent Gaussians decays to zero. For $d>2$ the analysis is
complex and depends on the transient behavior of the mean values
and variances (and not only on their steady state values), but a
reasonable conjecture is to assume that a same form of criterion
is also minimized for $d>2$. The result is different from the ML
lattice point, which minimizes $\left\| \boldsymbol{G} \cdot
\underline{b} - \underline{y}\right\|^2$, where both criteria give
the same point for weak noise but may give different solutions for
strong noise. This may explain the experiments where the iterative
decoder is successful in decoding the ML point for the AWGN
channel near channel capacity, but fails in quantization or
shaping applications where the effective noise is much stronger.
These results also show that the iterative decoder converges to
impulses at the coordinates of a single lattice point. It was then
explained that analyzing the amplitudes of consistent Gaussians is
not sufficient, so these results can not be regarded as a complete
convergence analysis. However, the analysis gave a set of
necessary conditions on $\boldsymbol{H}$, and also led to useful
insights to the convergence process.

\section {Code Design}
\label{code_design}
%Since we still lack full understanding of the dependency of LDLC error spectrum on the code parameters, we shall use a combination of intuitive arguments and simulation trial and error.
%Simulations show no reason to prefer irregular LDLC over regular ones. In addition, we shall concentrate on magic square LDLC, since as explained in section \ref{qual_exp} and demonstrated in figure \ref{check_waveforms}, it is beneficial to have nonzero elements with various magnitudes at each row and column.
\subsection{Choosing the Generating Sequence} \label{choose_seq}
We shall concentrate on magic square LDLC, since they have inherent diversity of the nonzero elements in each row and column, which was shown above to be beneficial.
It still remains to choose the LDLC generating sequence $h_1, h_2,...h_d$. Assume that the algorithm converged, and each PDF has a peak at the desired value. When the periodic functions are multiplied at a variable node, the correct peaks will then align. We would like that all the other peaks will be strongly attenuated, i.e. there will be no other point where the peaks align. This resembles the definition of the least common multiple (LCM) of integers: if the periods were integers, we would like to have their LCM as large as possible. This argument suggests the sequence $\{1/2, 1/3, 1/5, 1/7, 1/11, 1/13, 1/17,...\}$, i.e. the reciprocals of the smallest $d$ prime numbers. Since the periods are $1/h_1, 1/h_2,...1/h_d$, we will get the desired property. Simulations have shown that increasing $d$ beyond $7$ with this choice gave negligible improvement. % (the added numbers are already very small and have almost no effect).
Also, performance was improved by adding some ``dither'' to the sequence, resulting in $\{1/2.31, 1/3.17, 1/5.11, 1/7.33, 1/11.71, 1/13.11, 1/17.55\}$. For $d<7$, the first $d$ elements are used.

An alternative approach is a sequence of the form $\{1, \epsilon, \epsilon,...,\epsilon\}$, where $\epsilon << 1$. For this case, every variable node will receive a single message with period $1$ and $d-1$ messages with period $1/\epsilon$. For small $\epsilon$, the period of these $d-1$ messages will be large and multiplication by the channel Gaussian will attenuate all the unwanted replicas. The single remaining replica will attenuate all the unwanted replicas of the message with period 1. A convenient choice is $\epsilon=\frac{1}{\sqrt{d}}$, which ensures that $\alpha = \frac{d-1}{d} < 1$, as required by Theorem \ref{var_behavior}. As an example, for $d=7$ the sequence will be $\{1, \frac{1}{\sqrt{7}}, \frac{1}{\sqrt{7}}, \frac{1}{\sqrt{7}}, \frac{1}{\sqrt{7}}, \frac{1}{\sqrt{7}}, \frac{1}{\sqrt{7}}\}$.

\subsection{Necessary Conditions on $\boldsymbol{H}$} \label{nec_cond}
The magic square LDLC definition and convergence analysis imply four necessary conditions on $\boldsymbol{H}$:
\begin{enumerate}
\item $|det(\boldsymbol{H})|=1$. This condition is part of the LDLC definition, which ensures proper density of the lattice points in $\mathbb{R}^m$. If $|det(\boldsymbol{H})| \neq 1$, it can be easily normalized by dividing $\boldsymbol{H}$ by $\sqrt[n]{|det(\boldsymbol{H})|}$. Note that practically we can allow $|det(\boldsymbol{H})| \neq 1$ as long as $\sqrt[n]{|det(\boldsymbol{H})|} \approx 1$, since $\sqrt[n]{|det(\boldsymbol{H})|}$ is the gain factor of the transmitted codeword. For example, if $n=1000$, having $|det(\boldsymbol{H})| = 0.01$ is acceptable, since we have $\sqrt[n]{|det(\boldsymbol{H})|} = 0.995$, which means that the codeword has to be further amplified by $20 \cdot \log_{10}(0.995) = 0.04$ dB, which is negligible.

Note that normalizing $\boldsymbol{H}$ is applicable only if $\boldsymbol{H}$ is non-singular. If $\boldsymbol{H}$ is singular, a row and a column should be sequentially omitted until $\boldsymbol{H}$ becomes full rank. This process may result in slightly reducing $n$ and a slightly different row and column degrees than originally planned.
\item $\alpha < 1$, where
$\alpha \stackrel{\Delta}{=} \frac{\sum_{i=2}^d h_i^2}{h_1^2}$.
This guarantees exponential convergence rate for the variances
(Theorem \ref{var_behavior}). Choosing a smaller $\alpha$ results in faster convergence, but we should not take
$\alpha$ too small since the steady state variance of the wide
variable node messages, as well as the steady state error of the
mean values of these messages, increases when $\alpha$ decreases,
as discussed in Section \ref{mean_convergence}. This may result in
deviation of the decoded codeword from the ML codeword, as
discussed in Section \ref{amp_convergence}. For the first LDLC
generating sequence of the previous subsection, we have $\alpha =
0.92$ and $0.87$ for $d=7$ and $5$, respectively, which is a
reasonable trade off. For the second sequence type we have $\alpha
= \frac{d-1}{d}$.
\item All the eigenvalues of $\tilde{\boldsymbol{H}}$ must have magnitude less than $1$, where $\tilde{\boldsymbol{H}}$ is defined in Theorem \ref{narrow_mean_theorem}. This is a necessary condition for convergence of the mean values of the narrow messages. Note that adding random signs to the nonzero $\boldsymbol{H}$ elements is essential to fulfill this necessary condition, as explained in Section \ref{mean_convergence}.
\item All the eigenvalues of $\boldsymbol{F}$ must have magnitude less than $1$, where $\boldsymbol{F}$ is defined in Theorem \ref{wide_mean_theorem}. This is a necessary condition for convergence of the mean values of the wide messages.
\end{enumerate}
Interestingly, it comes out experimentally that for large codeword length $n$ and relatively small degree $d$ (e.g. $n \geq 1000 $ and $d \leq 10$), a magic square LDLC with generating sequence that satisfies $h_1 = 1$ and $\alpha < 1$ results in $\boldsymbol{H}$ that satisfies all these four conditions: $\boldsymbol{H}$ is nonsingular without any need to omit rows and columns, $\sqrt[n]{|det(\boldsymbol{H})|} \approx 1$ without any need for normalization, and all eigenvalues of $\tilde{\boldsymbol{H}}$ and $\boldsymbol{F}$ have magnitude less than 1 (typically, the largest eigenvalue of $\tilde{\boldsymbol{H}}$ or $\boldsymbol{F}$ has magnitude of $0.94 - 0.97$, almost independently of $n$ and the choice of nonzero $\boldsymbol{H}$ locations). Therefore, by simply dividing the first generating sequence of the previous subsection by its first element, the constructed $\boldsymbol{H}$ meets all the necessary conditions, where the second type of sequence meets the conditions without any need for modifications.

\subsection{Construction of $\boldsymbol{H}$ for Magic Square LDLC} \label{loop_def}
We shall now present a simple algorithm for constructing a parity check matrix for a magic square LDLC. If we look at the bipartite graph, each variable node and each check node has $d$ edges connected to it, one with every possible weight $h_1, h_2,...h_d$. All the edges that have the same weight $h_j$ form a permutation from the variable nodes to the check nodes (or vice versa). The proposed algorithm generates $d$ random permutations and then searches sequentially and cyclically for 2-loops (two parallel edges from a variable node to a check node) and 4-loops (two variable nodes that both are connected to a pair of check nodes). When such a loop is found, a pair is swapped in one of the permutations such that the loop is removed. A detailed pseudo-code for this algorithm is given in Appendix \ref{LDLC_gen_app}.

\section {Decoder Implementation} %considerations}
\label{implementation}
%\subsection{Resolution and range of the PDFs}
Each PDF should be approximated with a discrete vector with resolution $\Delta$ and finite range. According to the Gaussian Q-function, choosing a range of, say, $6\sigma$ to both sides of the noisy channel observation will ensure that the error probability due to PDF truncation will be $ \approx 10^{-9}$. Near capacity, $\sigma^2 \approx \frac{1}{2 \pi e}$, so $12\sigma \approx 3$.
Simulation showed that resolution errors became negligible for $\Delta = 1/64$. Each PDF was then stored in a $L=256$ elements vector, corresponding to a range of size $4$.

At the check node, the PDF $f_j(x)$ that arrives from variable node $j$ is first expanded by $h_j$ (the appropriate coefficient of $\boldsymbol{H}$) to get $f_j(x/h_j)$. In a discrete implementation with resolution $\Delta$ the PDF is a vector of values $f_j(k\Delta)$, $k \in \mathbb{Z}$. As described in Section \ref{code_design}, we shall usually use $h_j \leq 1$ so the expanded PDF will be shorter than the original PDF. If the expand factor $1/|h_j|$ was an integer, we could simply decimate $f_j(k\Delta)$ by $1/|h_j|$. However, in general it is not an integer so we should use some kind of interpolation. The PDF $f_j(x)$ is certainly not band limited, and as the iterations go on it approaches an impulse, so simple interpolation methods (e.g. linear) are not suitable. Suppose that we need to calculate $f_j((k+\epsilon)\Delta)$, where $-0.5 \leq \epsilon \leq 0.5$. A simple interpolation method which showed to be effective is to average $f_j(x)$ around the desired point, where the averaging window length $l_w$ is chosen to ensure that every sample of $f_j(x)$ is used in the interpolation of at least one output point. This ensures that an impulse can not be missed. The interpolation result is then $\frac{1}{2l_w+1}\sum_{i=-l_w}^{l_w} f_j((k-i)\Delta)$, where $l_w=\left\lfloor \frac {\left\lceil 1/|h_j|\right\rceil}{2}\right\rfloor$.

%\subsection{Convolution and periodic extension}
The most computationally extensive step at the check nodes is the calculation the convolution of $d-1$ expanded PDF's. An efficient method is to calculate the fast Fourier transforms (FFTs) of all the PDF's, multiply the results and then perform inverse FFT (IFFT). The resolution of the FFT should be larger than the expected convolution length, which is roughly $L_{out} \approx L \cdot \sum_{i=1}^d h_i$, where $L$ denotes the original PDF length. Appendix \ref{fft_app} shows a way to use FFTs of size $1/\Delta$, where $\Delta$ is the resolution of the PDF. Usually $1/\Delta << L_{out}$ so FFT complexity is significantly reduced. Practical values are $L=256$ and $\Delta=1/64$, which give an improvement factor of at least $4$ in complexity.

%Note that in section \ref{decoder}, an alternative approach was suggested for the final decision step of the decoder. In this approach, 
%The stretching of PDF's is done using interpolation, that averages the neighboring points of the desired location. % in a manner that all the PDF samples are used, such that an impulse will not be missed.
%Multiplication of check node messages is preceded by widening the messages by 1 sample to each side. %, in order to avoid impulses that do not overlap due to quantization effects.
%These operations inhibit impulses from disappearing due to finite resolution.
%\subsection{Multiplication at the variable nodes}
Each variable node receives $d$ check node messages. The output variable node message is calculated by generating the product of $d-1$ input messages and the channel Gaussian. As the iterations go on, the messages get narrow and may become impulses, with only a single nonzero sample. Quantization effects may cause impulses in two messages to be shifted by one sample. This will result in a zero output (instead of an impulse). Therefore, it was found useful to widen each check node message $Q(k)$ prior to multiplication, such that $Q_w(k) = \sum_{i=-1}^1 Q(k+i)$, i.e. the message is added to its right shifted and left shifted (by one sample) versions.

\section{Computational Complexity and Storage Requirements} \label{complexity}
Most of the computational effort is invested in the $d$ FFT's and $d$ IFFT's (of length $1/\Delta$) that each check node performs each iteration. The total number of multiplications for $t$ iterations is $o\left(n \cdot d \cdot t \cdot \frac{1}{\Delta} \cdot \log_2(\frac{1}{\Delta})\right)$.
%
%A reasonable measure of computational complexity for the iterative algorithm is the number of multiplications. Each check node performs $d$ FFT's and $d$ IFFT's of length $L/D = 1/\Delta$ at each iteration. Each FFT therefore requires $o\left(\frac{1}{\Delta} \cdot log_2(\frac{1}{\Delta})\right)$ multiplications. The check node also performs multiplication of the FFT results, but this requires only $1/\Delta$ multiplication per each FFT, so it can be neglected (as well as the interpolations). Therefore, the total number of multiplications is $o\left(n \cdot d \cdot t \cdot \frac{1}{\Delta} \cdot log_2(\frac{1}{\Delta})\right)$.
As in binary LDPC codes, the computational complexity has the attractive property of being linear with block length. However, the constant that precedes the linear term is significantly higher, mainly due to the FFT operations.

The memory requirements are governed by the storage of the $nd$ check node and variable node messages, with total memory of $o(n \cdot d \cdot L)$. Compared to binary LDPC, the factor of $L$ significantly increases the required memory. For example, for $n=10,000$, $d=7$ and $L=256$, the number of storage elements is of the order of $10^7$.

\section {Encoding and Shaping} \label{enc_shape}

\subsection{Encoding} \label{enc}
The LDLC encoder has to calculate $\underline{x} = \boldsymbol{G}
\cdot \underline{b}$, where $\underline{b}$ is an integer message
vector. Note that unlike $\boldsymbol{H}$,
$\boldsymbol{G}=\boldsymbol{H}^{-1}$ is not sparse, in general, so
the calculation requires computational complexity and storage of
$o(n^2)$. This is not a desirable property because the decoder's
computational complexity is only $o(n)$. A possible solution is to
use the Jacobi method \cite{sparse} to solve $\boldsymbol{H} \cdot
\underline{x} = \underline{b}$, which is a system of sparse linear
equations. Using this method, a magic square LDLC encoder
calculates several iterations of the form:
\begin{align}
\underline{x}^{(t)} = \underline{\tilde{b}} - \tilde{\boldsymbol{H}} \cdot \underline{x}^{(t-1)}
\end{align}
with initialization $\underline{x}^{(0)}=\underline{0}$. The matrix $\tilde{\boldsymbol{H}}$ is defined in Lemma \ref{narrow_mean_lemma} of Section \ref{mean_convergence}. The vector $\underline{\tilde{b}}$ is a permuted and scaled version of the integer vector $\underline{b}$, such that the $i$'th element of $\underline{\tilde{b}}$ equals the element of $\underline{b}$ for which the appropriate row of $\boldsymbol{H}$ has its largest magnitude value at the $i$'th location. This element is further divided by this largest magnitude element.

A necessary and sufficient condition for convergence to $\underline{x} = \boldsymbol{G} \cdot \underline{b}$ is that all the eigenvalues of $\tilde{\boldsymbol{H}}$ have magnitude less than $1$ \cite{sparse}. However, it was shown that this is also a necessary condition for convergence of the LDLC iterative decoder (see Sections \ref{mean_convergence}, \ref{nec_cond}), so it is guaranteed to be fulfilled for a properly designed magic square LDLC. Since $\tilde{\boldsymbol{H}}$ is sparse, this is an $o(n)$ algorithm, both in complexity and storage.

\subsection{Shaping} \label{shaping}
For practical use with the power constrained AWGN channel, the encoding operation must be accompanied by shaping, in order to prevent the transmitted codeword's power from being too large. Therefore, instead of mapping the information vector $\underline{b}$ to the lattice point $\underline{x} = \boldsymbol{G} \cdot \underline{b}$, it should be mapped to some other lattice point $\underline{x}' = \boldsymbol{G} \cdot \underline{b}'$, such that the lattice points that are used as codewords belong to a shaping region (e.g. an $n$-dimensional sphere). The shaping operation is the mapping of the integer vector $\underline{b}$ to the integer vector $\underline{b}'$.

As explained in Section \ref{lattice_codes}, this work concentrates on the lattice design and the lattice decoding algorithm, and not on the shaping region or shaping algorithms. Therefore, this section will only highlight some basic shaping principles and ideas.

A natural shaping scheme for lattice codes is nested lattice coding \cite{Zamir_Erez}. In this scheme, shaping is done by quantizing the lattice point $\boldsymbol{G} \cdot \underline{b}$ onto a coarse lattice $\boldsymbol{G}'$, where the transmitted codeword is the quantization error, which is uniformly distributed along the Voronoi cell of the coarse lattice. If the second moment of this Voronoi cell is close to that of an $n$-dimensional sphere, the scheme will attain close-to-optimal shaping gain. Specifically, assume that the information vector $\underline{b}$ assumes integer values in the range $0,1,...M-1$ for some constant integer $M$. Then, we can choose the coarse lattice to be $\boldsymbol{G'} = M\boldsymbol{G}$. The volume of the Voronoi cell for this lattice is $M^n$, since we assume $det(\boldsymbol{G})=1$ (see Section \ref{lattice_codes}). If the shape of the Voronoi cell resembles an $n$-dimensional sphere (as expected from a capacity approaching lattice code), it will attain optimal shaping gain (compared to uncoded transmission of the original integer sequence $\underline{b}$).

The shaping operation will find the coarse lattice point $M \boldsymbol{G} \underline{k}$, $\underline{k} \in \mathbb{Z}^n$, which is closest to the fine lattice point $\underline{x} = \boldsymbol{G} \cdot \underline{b}$. The transmitted codeword will be:
\begin{align*}
\underline{x}' = \underline{x}-M\boldsymbol{G}\underline{k}=\boldsymbol{G}(\underline{b}-M\underline{k})=\boldsymbol{G}\underline{b'}
\end{align*}
where $\underline{b}' \stackrel{\Delta}{=} \underline{b}-M\underline{k}$ (note that the ``inverse shaping'' at the decoder, i.e. transforming from $\underline{b}'$ to $\underline{b}$, is a simple modulo calculation: $\underline{b} = \underline{b}' \mod M$).
Finding the closest coarse lattice point $M \boldsymbol{G} \underline{k}$ to $\underline{x}$ is equivalent to finding the closest fine lattice point $\boldsymbol{G} \cdot \underline{k}$ to the vector $\underline{x}/M$. This is exactly the operation of the iterative LDLC decoder, so we could expect that is could be used for shaping. However, simulations show that the iterative decoding finds a vector $\underline{k}$ with poor shaping gain. The reason is that for shaping, the effective noise is much stronger than for decoding, and the iterative decoder fails to find the nearest lattice point if the noise is too large (see Section \ref{amp_convergence}).

Therefore, an alternative algorithm has to be used for finding the nearest coarse lattice point. The complexity of finding the nearest lattice point grows exponentially with the lattice dimension $n$ and is not feasible for large dimensions \cite{Agrell_lattice}. However, unlike decoding, for shaping applications it is not critical to find the exact nearest lattice point, and approximate algorithms may be considered (see \cite{signal}). A possible method \cite{Fano_lattice} is to perform QR decomposition on $\boldsymbol{G}$ in order to transform to a lattice with upper triangular generator matrix, and then use sequential decoding algorithms (such as the Fano algorithm) to search the resulting tree. The main disadvantage of this approach is computational complexity and storage of at least $o(n^2)$.
Finding an efficient shaping scheme for LDLC is certainly a topic for further research.

\section {Simulation Results}
\label{sim_res}
\begin{figure}
\centering
%\vspace {-0.3cm}
\includegraphics[width=3.5in]{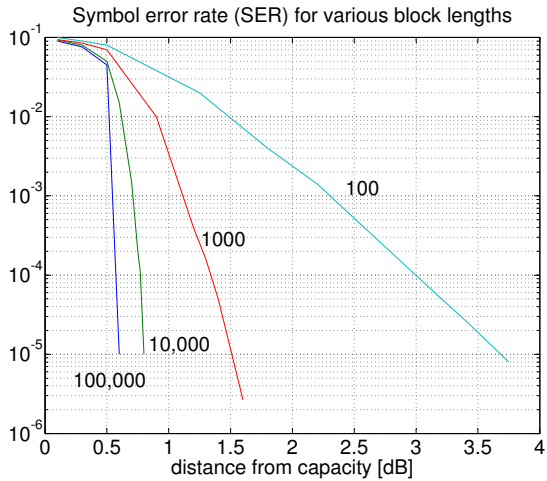}  %, width=2.5in [height=2in]
%\vspace {-0.5cm}
\caption{Simulation results}
%\vspace{-0.7cm}
\label{sim_res_plot}
\end{figure}

Magic square LDLC with the first generating sequence of Section \ref{choose_seq} (i.e. $\{1/2.31, 1/3.17, 1/5.11, 1/7.33, 1/11.71, 1/13.11, 1/17.55\}$) were simulated for the AWGN channel at various block lengths.
% $n$ of 100, 1000, 10000, and 100000. All simulations were for magic square LDLC with the sequence of nonzero elements that was described in section \ref{code_design}, i.e. $\{h_1, h_2,...h_7\} = 2.31 \cdot \{1/2.31, 1/3.17, 1/5.11, 1/7.33, 1/11.71, 1/13.11, 1/17.55\}$ (the multiplication by $2.31$ ensures that the largest element is $1$).
The degree was $d=5$ for $n=100$ and $d=7$ %and the first $5$ elements of the sequence were used.
for all other $n$. For $n=100$ the matrix $\boldsymbol{H}$ was further normalized to get $\sqrt[n]{det(\boldsymbol{H})}=1$. For all other $n$, normalizing the generating sequence such that the largest element has magnitude $1$ also gave the desired determinant normalization (see Section \ref{nec_cond}). The $\boldsymbol{H}$ matrices were generated using the algorithm of Section \ref{loop_def}. %, such that the resulting bipartite graph does not include 4-loops.
PDF resolution was set to $\Delta = 1/256$ with a total range of $4$, i.e. each PDF was represented by a vector of $L=1024$ elements.
%so the PDF vector length was $1024$. Note decreasing the resolution to $\Delta = 1/64$ resulted in minor degradation, but
High resolution was used since our main target is to prove the LDLC concept and eliminate degradation due to implementation considerations. For this reason, the decoder was used with 200 iterations (though most of the time, a much smaller number was sufficient).

%According to Poltyrev's results (section \ref{lattice_codes}), shaping is not necessary and
In all simulations the all-zero codeword was used. Approaching channel capacity is equivalent to $\sigma^2 \rightarrow \frac{1}{2 \pi e}$ (see Section \ref{lattice_codes}), so performance is measured in symbol error rate (SER), vs. the distance of the noise variance $\sigma^2$ from capacity (in dB).
%
%We shall not restrict ourselves to a specific rate (i.e. finite symbol constellation), but will use an infinite constellation. As described in section \ref{lattice_codes}, Poltyrev \cite{Poltyrev} has defined a generalized channel capacity for the AWGN without restriction and showed that when a power constraint is added and a spherical shaping region is used, this definition is coincides with the channel capacity of the power limited AWGN. For a lattice with determinant $1$,the generalized channel capacity is approached when $\sigma^2 \rightarrow \frac{1}{2 \pi e}$. The results below measure SER as a function of the distance of the noise variance $\sigma^2$ from capacity (in dB), which is defined as $10 \cdot log_{10} (2 \pi e \sigma^2)$.
%
%As described in the introduction, this work concentrates on the lattice design, and the issue of practical shaping or encoding algorithms is not addressed. However, due to the linearity of the lattice, the error performance does not depend on the transmitted codeword. Therefore, all the simulation measure SER when the all-zero codeword is transmitted, so encoding is avoided.
The results are shown in Figure \ref{sim_res_plot}. %It can be seen that performance improves as the block length is increased, as expected.
At SER of $10^{-5}$, for $n=100000$, $10000$, $1000$, $100$ we can work as close as $0.6$dB, $0.8$dB, $1.5$dB and $3.7$dB from capacity, respectively.

Similar results were obtained for $d=7$ with the second type of generating sequence of Section \ref{choose_seq}, i.e. $\{1, \frac{1}{\sqrt{7}}, \frac{1}{\sqrt{7}}, \frac{1}{\sqrt{7}}, \frac{1}{\sqrt{7}}, \frac{1}{\sqrt{7}}, \frac{1}{\sqrt{7}}\}$. Results were slightly worse than for the first generating sequence (by less than 0.1 dB). Increasing $d$ did not give any visible improvement.

\section {Conclusion}
Low density lattice codes (LDLC) were introduced. LDLC are novel lattice codes that can
approach capacity and be decoded efficiently. Good error performance within $\sim 0.5$dB from capacity at block length of 100,000 symbols was demonstrated. Convergence analysis was presented for the iterative decoder, which is not complete, but yields necessary conditions on $\boldsymbol{H}$ and significant insight to the convergence process. Code parameters were chosen from intuitive arguments, so it is reasonable to assume that when the code structure will be more understood, better parameters could be found, and channel capacity could be approached even closer.

Multi-input, multi-output (MIMO) communication systems have become popular in recent years. Lattice codes have been proposed in this context as space-time codes (LAST) \cite{LAST}. The concatenation of the lattice encoder and the MIMO channel generates a lattice. If LDLC are used as lattice codes and the MIMO configuration is small, the inverse generator matrix of this concatenated lattice can be assumed to be sparse. Therefore, the MIMO channel and the LDLC can be jointly decoded using an LDLC-like decoder. However, even if a magic square LDLC is used as the lattice code, the concatenated lattice is not guaranteed to be equivalent to a magic square LDLC, and the necessary conditions for convergence are not guaranteed to be fulfilled. Therefore, the usage of LDLC for MIMO systems is a topic for further research.

\appendices
\section{Exact PDF Calculations} \label{app_exact_pdf}
Given the $n$-dimensional noisy observation $\underline{y}=\underline{x}+\underline{w}$ of the transmitted codeword $\underline{x}=\boldsymbol{G}\underline{b}$, we would like to calculate the probability density function (PDF)  $f_{x_k|\underline{y}}(x_k|\underline{y})$. We shall start by calculating 
$
f_{\underline{x}|\underline{y}}(\underline{x}|\underline{y}) = %\frac{f_{\underline{x},\underline{y}}(\underline{x},\underline{y})}{f_{\underline{y}}(\underline{y})} =
\frac{f_{\underline{x}}(\underline{x})f_{\underline{y}|\underline{x}}(\underline{y}|\underline{x})}{f_{\underline{y}}(\underline{y})}
$. Denote the shaping region by $B$ ($\boldsymbol{G}$ will be used to denote both the lattice and its generator matrix).
$f_{\underline{x}}(\underline{x})$ is a sum of $|\boldsymbol{G} \cap B|$ $n$-dimensional Dirac delta functions, since $\underline{x}$ has nonzero probability only for the lattice points that lie inside the shaping region. Assuming further that all codewords are used with equal probability, all these delta functions have equal weight of $\frac{1}{|\boldsymbol{G} \cap B|}$.
The expression for $f_{\underline{y}|\underline{x}}(\underline{y}|\underline{x})$ is simply the PDF of the i.i.d Gaussian noise vector. We therefore get:
\begin{align} \label{direct_delta_calc}
f_{\underline{x}|\underline{y}}(\underline{x}|\underline{y}) =
\frac{f_{\underline{x}}(\underline{x})f_{\underline{y}|\underline{x}}(\underline{y}|\underline{x})}{f_{\underline{y}}(\underline{y})} =
\end{align}
%\nonumber\\
\begin{align*}
\frac{\frac{1}{|\boldsymbol{G} \cap B|} \sum_{\underline{l} \in \boldsymbol{G} \cap B} \delta (\underline{x} - \underline{l}) \cdot (2\pi\sigma^2)^{-n/2} e^{-\sum_{i=1}^{n}(y_i - x_i)^2/2\sigma^2}}{f_{\underline{y}}(\underline{y})} =
\end{align*}
%\nonumber\\
\begin{align*}
= C \cdot \sum_{\underline{l} \in \boldsymbol{G} \cap B} \delta (\underline{x} - \underline{l}) \cdot e^{-d^2(\underline{l}, \underline{y})/2\sigma^2}
\end{align*}
Where $C$ is not a function of $\underline{x}$ and $d^2(\underline{l}, \underline{y})$ is the squared Euclidean distance between the vectors $\underline{l}$ and $\underline{y}$ in $\mathbb{R}^n$. It can be seen that the conditional PDF of $\underline{x}$ has a delta function for each lattice point, located at this lattice point with weight that is proportional to the exponent of the negated squared Euclidean distance of this lattice point from the noisy observation. The ML point corresponds to the delta function with largest weight.

As the next step, instead of calculating the $n$-dimensional PDF of the whole vector $\underline{x}$, we shall calculate the $n$ one-dimensional PDF's for each of the components $x_k$ of the vector $\underline{x}$ (conditioned on the whole observation vector $\underline{y}$):
\begin{align}
f_{x_k|\underline{y}}(x_k|\underline{y}) =
\end{align}
\begin{align*}
\int\int_{x_i, i \neq k}\cdots\int f_{\underline{x}|\underline{y}}(\underline{x}|\underline{y}) d x_1 d x_2 \cdots d x_{k-1} d x_{k+1} \cdots d x_n =
\end{align*}
\begin{align*}
= C \cdot \sum_{\underline{l} \in \boldsymbol{G} \cap B} \delta (x_k - l_k) \cdot e^{-d^2(\underline{l}, \underline{y})/2\sigma^2}
\end{align*}

This finishes the proof of (\ref{exact_solution}).
It can be seen that the conditional PDF of $x_k$ has a delta function for each lattice point, located at the projection of this lattice point on the coordinate $x_k$, with weight that is proportional to the exponent of the negated squared Euclidean distance of this lattice point from the noisy observation. The ML point will therefore correspond to the delta function with largest weight in each coordinate. Note, however, that if several lattice points have the same projection on a specific coordinate, the weights of the corresponding delta functions will add and may exceed the weight of the ML point.

\section{Extending Gallager's Technique to the Continuous Case} \label{trick_pdf}
In \cite{Gallager}, the derivation of the LDPC iterative decoder was simplified using the following technique: the codeword elements $x_k$ were assumed i.i.d. and a condition was added to all the probability calculations, such that only valid codewords were actually considered. The question is then how to choose the marginal PDF of the codeword elements. In \cite{Gallager}, binary codewords were considered, and the i.i.d distribution assumed the values '0' and '1' with equal probability. Since we extend the technique to the continuous case, we have to set the continuous marginal distribution $f_{x_k}(x_k)$. It should be set such that $f_{\underline{x}}(\underline{x})$, assuming that $\underline{x}$ is a lattice point, is the same as $f(\underline{x}|\underline{s} \in \mathbb{Z}^n)$, assuming that $x
_k$ are i.i.d with marginal PDF $f_{x_k}(x_k)$, where $\underline{s}
\stackrel{\Delta}{=}\boldsymbol{H} \cdot \underline{x}$. This $f_{\underline{x}}(\underline{x})$ equals a weighted sum of Dirac delta functions at all lattice points, where the weight at each lattice point equals the probability to use this point as a codeword.

Before proceeding, we need the following property of conditional probabilities. For any two continuous valued RV's $u$, $v$ we have:
\begin{eqnarray} \label{conditional_pdf_property}
f(u|v \in \{v_1, v_2, ..., v_N\}) = \frac{\sum_{k=1}^N f_{u,v}(u,v_k)}{\sum_{k=1}^N f_v(v_k)}
% = \frac{\sum_{k=1}^N f_v(v_k)f(u|v_k)}{\sum_{k=1}^N f_v(v_k)}
\end{eqnarray}
(This property can be easily proved by following the lines of \cite{Papoulis}, pp. 159-160, and can also be extended to the infinite sum case).

Using (\ref{conditional_pdf_property}), we now have:
\begin{align*}
f(\underline{x}|\underline{s} \in \mathbb{Z}^n) = \frac{\sum_{\underline{i} \in \mathbb{Z}^n} f_{\underline{x}, \underline{s}}(\underline{x}, \underline{s} =
\underline{i})}{\sum_{\underline{i} \in \mathbb{Z}^n} f_{\underline{s}}(\underline{i})} =
\end{align*}
\begin{align}
=C \sum_{\underline{i} \in \mathbb{Z}^n} f(\underline{x}) f(\underline{s} = \underline{i}|\underline{x}) = C' \sum_{\underline{i} \in \mathbb{Z}^n} f(\underline{x}) \delta(\underline{x}-\boldsymbol{G} \underline{i})
\end{align}
where $C, C'$ are independent of $\underline{x}$.

The result is a weighted sum of Dirac delta functions at all lattice points, as desired. Now, the weight at each lattice point should equal the probability to use this point as a codeword. Therefore, $f_{x_k}(x_k)$ should be chosen such that at each lattice point, the resulting vector distribution $f_{\underline{x}}(\underline{x})=\prod_{k=1}^n f_{x_k}(x_k)$ will have a value that is proportional to the probability to use this lattice point. At $\underline{x}$ which is not a lattice point, the value of $f_{\underline{x}}(\underline{x})$ is not important.

\section{Derivation of the Iterative Decoder}
\label{iter_deriv}
In this appendix we shall derive the LDLC iterative decoder for a code with dimension $n$, using the tree assumption and Gallager's trick. 

Referring to figure \ref{tier_diagram},
assume that there are only 2 tiers. Using Gallager's trick we assume that the $x_k$'s are i.i.d. We would like to calculate $f(x_1|(\underline{y}, \underline{s} \in \mathbb{Z}^n)$, where $\underline{s}
\stackrel{\Delta}{=}\boldsymbol{H} \cdot \underline{x}$.  %(the i.i.d superscript should appear on all PDF's but it is omitted for convenience. We also omit the subscript of the PDF's, e.g. $f_x(x)$ is denoted by $f(x)$).
Due to the tree assumption, we can do it in two steps:

1. calculate the conditional PDF of the tier 1 variables of $x_1$, conditioned only on the check equations that relate the tier 1 and tier 2 variables.

2. calculate the conditional PDF of $x_1$ itself, conditioned only on the check equations that relate $x_1$ and its first tier variables, but using the results of step 1 as the PDF's for the tier 1 variables. Hence, the results will be equivalent to conditioning on all the check equations.

There is a basic difference between the calculation in step 1 and step 2: the condition in step 2 involves all the check equations that are related to $x_1$, where in step 1 a single check equation is always omitted (the one that relates the relevant tier 1 element with $x_1$ itself).

Assume now that there are many tiers, where each tier contains distinct elements of $\underline{x}$ (i.e. each element appears only once in the resulting tree). We can then start at the farthest tier and start moving toward $x_1$. We do it by repeatedly calculating step 1. After reaching tier 1, we use step 2 to finally calculate the desired conditional PDF for $x_1$.

This approach suggests an iterative algorithm for the calculation of $f(x_k|(\underline{y}, \underline{s} \in \mathbb{Z}^n)$ for $k$=$1,2..n$. In this approach we assume that the resulting tier diagram for each $x_k$ contains distinct elements for several tiers (larger or equal to the number of required iterations). We then repeat step 1 several times, where the results of the previous iteration are used as initial PDF's for the next iteration. Finally, we perform step 2 to calculate the final results.

Note that by conditioning only on part of the check equations in each iteration, we can not restrict the result to the shaping region. This is the reason that the decoder performs lattice decoding and not exact ML decoding, as described in Section \ref{iter_dec_sec}.
%is  $B$ (i.e. $\underline{s} \in i_B$) and must relax the condition, and only demand that the result will be a lattice point (i.e. $\underline{s} \in \mathbb{Z}^n$). This approximate decoding method (decoding to the infinite lattice and ignoring the boundaries) is no longer exact maximum likelihood decoding, and is usually denoted ``lattice decoding''.

We shall now turn to derive the basic iteration of the algorithm. For simplicity, we shall start with the final step of the algorithm (denoted step 2 above). We would like to perform $t$ iterations, so assume that for each $x_k$ there are $t$ tiers with a total of $N_c$ check equations. For every $x_k$ we need to calculate $f(x_k| \underline{s} \in \mathbb{Z}^{N_c},\underline{y}) = f(x_k| \underline{s}^{(tier_1)} \in \mathbb{Z}^{c_k}, \underline{s}^{(tier_2:tier_t)} \in \mathbb{Z}^{N_c-c_k},\underline{y})$, where $c_k$ is the number of check equations that involve $x_k$. $\underline{s}^{(tier_1)}\stackrel{\Delta}{=}\boldsymbol{H}^{(tier_1)} \cdot \underline{x}$ denotes the value of the left hand side of these check equations when $\underline{x}$ is substituted ($\boldsymbol{H}^{(tier_1)}$ is a submatrix of $\boldsymbol{H}$ that contains only the rows that relate to these check equations), and $\underline{s}^{(tier_2:tier_t)}$ relates in the same manner to all the other check equations. For simplicity of notations, denote the event $\underline{s}^{(tier_2:tier_t)} \in \mathbb{Z}^{N_c-c_k}$ by $A$. As explained above, in all the calculations we assume that all the $x_k$'s are independent.

Using (\ref{conditional_pdf_property}), we get:
\begin{eqnarray} \label{basic_eq}
f(x_k| \underline{s}^{(tier_1)} \in \mathbb{Z}^{c_k}, A, \underline{y}) =
\nonumber\\
=\frac{\sum_{\underline{i} \in \mathbb{Z}^{c_k}} f(x_k, \underline{s}^{(tier_1)}=\underline{i}|A, \underline{y})}{\sum_{\underline{i} \in \mathbb{Z}^{c_k}} f(\underline{s}^{(tier_1)}=\underline{i}|A, \underline{y})}
\end{eqnarray}
Evaluating the term inside the sum of the nominator, we get:
\begin{eqnarray} \label{denom}
f(x_k,\underline{s}^{(tier_1)}=\underline{i}|A,\underline{y}) =
\nonumber\\
= f(x_k|A,\underline{y}) \cdot f(\underline{s}^{(tier_1)}=\underline{i}|x_k, A, \underline{y})
\end{eqnarray}
Evaluating the left term, we get:
\begin{eqnarray} \label{left_denom}
f(x_k|A, \underline{y}) = f(x_k|y_k)
= \frac{f(x_k) f(y_k|x_k)}{f(y_k)} =
\nonumber\\
=\frac{f(x_k)}{f(y_k)} \cdot \frac{1}{\sqrt{2 \pi \sigma^2}} e^{-\frac{(y_k-x_k)^2}{2\sigma^2}}
\end{eqnarray}
where $f(x_k|\underline{y}) = f(x_k|y_k)$ due to the i.i.d assumption.
Evaluating now the right term of (\ref{denom}), we get:
\begin{eqnarray} \label{right_denom}
f(\underline{s}^{(tier_1)}=\underline{i}|x_k, A, \underline{y}) =
\nonumber\\
= \prod_{m=1}^{c_k} f(s_m^{(tier_1)} = i_m | x_k, A, \underline{y})
\end{eqnarray}
where $s_m^{(tier_1)}$ denotes the $m$'th component of $\underline{s}^{(tier_1)}$ and $i_m$ denotes the $m$'th component of $\underline{i}$. Note that each element of $\underline{s}^{(tier_1)}$ is a linear combination of several elements of $\underline{x}$. Due to the tree assumption, two such linear combinations have no common elements, except for $x_k$ itself, which appears in all linear combinations. However, $x_k$ is given, so the i.i.d assumption implies that all these linear combinations are independent, so (\ref{right_denom}) is justified. The condition $A$ (i.e.  $\underline{s}^{(tier_2:tier_t)} \in \mathbb{Z}^{N_c-c_k}$) does not impact the independence due to the tree assumption.

Substituting (\ref{denom}), (\ref{left_denom}), (\ref{right_denom}) back in (\ref{basic_eq}), we get:
\begin{align} \label{prod_sum_eq}
f(x_k| \underline{s}^{(tier_1)} \in \mathbb{Z}^{c_k}, A, \underline{y}) =
\end{align}
\begin{align*}
= C \cdot f(x_k) \cdot e^{-\frac{(y_k-x_k)^2}{2\sigma^2}} \sum_{\underline{i} \in \mathbb{Z}^{c_k}} \prod_{m=1}^{c_k} f(s_m^{(tier_1)} = i_m | x_k, A, \underline{y}) =
\end{align*}
\begin{align*}
= C \cdot f(x_k) \cdot e^{-\frac{(y_k-x_k)^2}{2\sigma^2}}
\sum_{i_1 \in \mathbb{Z}} \sum_{i_2 \in \mathbb{Z}} \cdots
\end{align*}
\begin{align*}
\cdots \sum_{i_{c_k} \in \mathbb{Z}}
\prod_{m=1}^{c_k} f(s_m^{(tier_1)} = i_m | x_k, A, \underline{y}) =
\end{align*}
\begin{align*}
= C \cdot f(x_k) \cdot e^{-\frac{(y_k-x_k)^2}{2\sigma^2}} \prod_{m=1}^{c_k} \sum_{i_m \in \mathbb{Z}} f(s_m^{(tier_1)} = i_m | x_k, A, \underline{y})
\end{align*}
where $C$ is independent of $x_k$.

We shall now examine the term inside the sum: $f(s_m^{(tier_1)} = i_m | x_k, A, \underline{y})$. Denote the linear combination that $s_m^{(tier_1)}$ represents by:
\begin{eqnarray}
s_m^{(tier_1)} = h_{m,1} x_k + \sum_{l=2}^{r_m}h_{m,l} x_{j_l}
\end{eqnarray}
where $\{h_{m,l}\}$, $l=1,2...r_m$ is the set of nonzero coefficients of the appropriate parity check equation, and ${j_l}$ is the set of indices of the appropriate $\underline{x}$ elements (note that the set ${j_l}$ depends on $m$ but we omit the ``$m$'' index for clarity of notations). Without loss of generality, $h_{m,1}$ is assumed to be the coefficient of $x_k$. Define $z_m \stackrel{\Delta}{=} \sum_{l=2}^{r_m}h_{m,l} x_{j_l}$, such that $s_m^{(tier_1)} = h_{m,1} x_k + z_m$. We then have:
\begin{align} \label{f_z_def}
f(s_m^{(tier_1)} = i_m | x_k, A, \underline{y}) =
\end{align}
\begin{align*}
=f_{z_m|x_k, A, \underline{y}}(z_m = i_m - h_{m,1} x_k | x_k, A, \underline{y})
\end{align*}
Now, since we assume that the elements of $\underline{x}$ are independent, the PDF of the linear combination $z_m$ equals the convolution of the PDF's of its components:
\begin{eqnarray} \label{conv_eq}
f_{z_m|x_k, A, \underline{y}}(z_m|x_k,A,\underline{y}) =
\nonumber\\ =\frac{1}{|h_{m,2}|}f_{x_{j_2}|A, \underline{y}}\left(\frac{z_m}{h_{m,2}}|A,\underline{y}\right) \circledast
\nonumber\\
\circledast \frac{1}{|h_{m,3}|}f_{x_{j_3}|A, \underline{y}}\left(\frac{z_m}{h_{m,3}}|A,\underline{y}\right) \circledast
\nonumber\\
\cdots \circledast \frac{1}{|h_{m,r_m}|}f_{x_{j_{r_m}}|A, \underline{y}}\left(\frac{z_m}{h_{m,r_m}}|A,\underline{y}\right)
\end{eqnarray}
Note that the functions $f_{x_{j_i}|\underline{y}}\left(x_{j_i}|A,\underline{y}\right)$ are simply the output PDF's of the previous iteration.

Define now
\begin{eqnarray} \label{p_def}
p_m(x_k) \stackrel{\Delta}{=}
f_{z_m|x_k, A,\underline{y}}(z_m=-h_{m,1} x_k|x_k,A,\underline{y})
\end{eqnarray}
Substituting (\ref{f_z_def}), (\ref{p_def}) in (\ref{prod_sum_eq}), we finally get:
\begin{align} \label{final_step2_eq}
f(x_k| \underline{s}^{(tier_1)} \in \mathbb{Z}^{c_k}, A, \underline{y}) =
\end{align}
\begin{align*}
= C \cdot f(x_k) \cdot e^{-\frac{(y_k-x_k)^2}{2\sigma^2}} \prod_{m=1}^{c_k} \sum_{i_m \in \mathbb{Z}} p_m(x_k - \frac{i_m}{h_{m,1}})
\end{align*}
%Note that $p_m(x)$ can be calculated by stretching the convolution result (\ref{conv_eq}) by a factor of $-h_{m,1}$.

This result can be summarized as follows. For each of the $c_k$ check equations that involve $x_k$, the PDF's (previous iteration results) of the active equation elements, except for $x_k$ itself, are expanded and convolved, according to (\ref{conv_eq}). The convolution result is scaled by $(-h_{m,1})$, the negated coefficient of $x_k$ in this check equation, according to (\ref{p_def}), to yield $p_m(x_k)$. Then, a periodic function with period $1/|h_{m,1}|$ is generated by adding an infinite number of shifted versions of the scaled convolution result, according to the sum term in (\ref{final_step2_eq}). After repeating this process for all the $c_k$ check equations that involve $x_k$, we get $c_k$ periodic functions, with possibly different periods. We then multiply all these functions. The multiplication result is further multiplied by the channel Gaussian PDF term $e^{-\frac{(y_k-x_k)^2}{2\sigma^2}}$ and finally by $f(x_k)$, the marginal PDF of $x_k$ under the i.i.d assumption. As discussed in Section \ref{iter_dec_sec}, we assume that $f(x_k)$ is a uniform distribution with large enough range. This means that $f(x_k)$ is constant over the valid range of $x_k$, and can therefore be omitted from (\ref{final_step2_eq}) and absorbed in the constant $C$.

%The term $f(x_k)$ is the marginal PDF of the component $x_k$ under the assumption that $\underline{x}$ is uniformly distributed in the shaping region $B$. If $B$ is an $n$-dimensional cube, then $f(x_k)$ will be a uniformly distributed PDF. In this case, $f(x_k)$ can be omitted from (\ref{final_step2_eq}) and absorbed in the constant $C$ (except for setting the interval for which the PDF is nonzero). If $B$ is a sphere,  $f(x_k)$ will approach a Gaussian distribution (for large $n$). If the variance of $x_k$ is significantly larger then the noise variance (i.e. high SNR) then the Gaussian noise PDF term will be much narrower than $f(x_k)$, and $f(x_k)$ can then be approximated by a constant (over the range for which the noise Gaussian in effectively nonzero), and omitted from (\ref{final_step2_eq}). Practically, we shall omit $f(x_k)$ from equation (\ref{final_step2_eq}), even for low SNR.

As noted above, this result is for the final step (equivalent to step 2 above), where we determine the PDF of $x_k$ according to the PDF's of all its tier 1 elements. However, the repeated iteration step is equivalent to step 1 above. In this step ,we assume that $x_k$ is a tier 1 element of another element, say $x_l$, and derive the PDF of $x_k$ that should be used as input to step 2 of $x_l$ (see figure \ref{tier_diagram}). It can be seen that the only difference between step 2 and step 1 is that in step 2 all the check equations that involve $x_k$ are used, where in step 1 the check equation that involves both $x_k$ and $x_l$ is ignored (there must be such an equation since $x_k$ is one of the tier 1 elements of $x_l$). Therefore, the step1 iteration is identical to (\ref{final_step2_eq}), except that the product does not contain the term that corresponds to the check equation that combines both $x_k$ and $x_l$. Denote
\begin{eqnarray}
f_{kl}(x_k) \stackrel{\Delta}{=} f(x_k| \underline{s}^{(tier_1\ except\ l)} \in \mathbb{Z}^{c_k-1}, A, \underline{y})
\end{eqnarray}
We then get:
\begin{eqnarray} \label{step1_eq}
f_{kl}(x_k) = C \cdot  e^{-\frac{(y_k-x_k)^2}{2\sigma^2}} \prod_{\substack{m=1 \\ m \neq m_l}}^{c_k} \sum_{i_m \in \mathbb{Z}} p_m(x_k - \frac{i_m}{h_{m,1}})
\end{eqnarray}
where $m_l$ is the index of the check equation that combines both $x_k$ and $x_l$. In principle, a different $f_{kl}(x_k)$ should be calculated for each $x_l$ for which $x_k$ is a tier 1 element. However, the calculation is the same for all $x_l$ that share the same check equation. Therefore, we should calculate $f_{kl}(x_k)$ once for each check equation that involves $x_k$. $l$ can be regarded as the index of the check equation within the set of check equations that involve $x_k$.

We can now formulate the iterative decoder. The decoder state variables are PDF's of the form $f^{(t)}_{kl}(x_k)$, where $k=1,2,...n$. For each $k$, $l$ assumes the values $1,2,...c_k$, where $c_k$ is the number of check equations that involve $x_k$. $t$ denotes the iteration index. For a regular LDLC with degree $d$ there will be $nd$ PDF's. The PDF's are initialized by assuming that $x_k$ is a leaf of the tier diagram. Such a leaf has no tier 1 elements, so $f_{kl}(x_k) =  f(x_k) \cdot f(y_k|x_k)$. As explained above for equation (\ref{final_step2_eq}), we shall omit the term $f(x_k)$, resulting in initialization with the channel noise Gaussian around the noisy observation $y_k$. Then, the PDF's are updated in each iteration according to (\ref{step1_eq}). The variable node messages should be further normalized in order to get actual PDF's, such that $\int_{-\infty}^{\infty}f_{kl}(x_k) dx_k = 1$ (this will compensate for the constant $C$). The final PDF's for $x_k$, $k=1,2,...n$ are then calculated according to (\ref{final_step2_eq}).

Finally, we have to estimate the integer valued information vector $\underline{b}$. This can be done by first estimating the codeword vector $\underline{x}$ from the peaks of the PDF's: $\hat{x_k} = arg \max_{x_k} f(x_k| \underline{s}^{(tier_1)} \in \mathbb{Z}^{c_k}, A, \underline{y})$. Finally, we can estimate $\underline{b}$ as $\underline{\hat{b}}=\left\lfloor \boldsymbol{H}\underline{\hat{x}}\right\rceil$.

We have finished developing the iterative algorithm. It can be easily seen that the message passing formulation of Section \ref{decoder} actually implements this algorithm.

\section{Asymptotic Behavior of the Variances Recursion}
\label{var_app}
\subsection{Proof of Lemma \ref{same_val_same_var} and Lemma \ref{var_recur_lemma}}
We shall now derive the basic iterative equations that relate the variances at iteration $t+1$ to the variances at iteration $t$ for a magic square LDLC with dimension $n$, degree $d$ and generating sequence $h_1 \geq h_2 \geq ...\geq h_d > 0$.

Each iteration, every check node generates $d$ output messages, one for each variable node that is connected to it, where the weights of these $d$ connections are $ \pm h_1, \pm h_2,...,\pm h_d$. For each such output message, the check node convolves $d-1$ expanded variable node PDF messages, and then stretches and periodically extends the result. For a specific check node, denote the variance of the variable node message that arrives along an edge with weight $\pm h_j$ by $V^{(t)}_j$, $j=1,2,...d$. Denote the variance of the message that is sent back to a variable node along an edge with weight $\pm h_j$ by $\tilde{V}^{(t)}_j$. From (\ref{iter_conv}), (\ref{check_message}), we get:
\begin{eqnarray} \label{var_check}
\tilde{V}^{(t)}_j = \frac{1}{h_j^2}\sum_{\substack{i=1 \\ i \neq j}}^d h_i^2 V^{(t)}_i
\end{eqnarray}
Then, each variable node generates $d$ messages, one for each check node that is connected to it, where the weights of these $d$ connections are $\pm h_1, \pm h_2,...,\pm h_d$. For each such output message, the variable node generates the product of $d-1$ check node messages and the channel noise PDF. For a specific variable node, denote the  variance of the message that is sent back to a check node along an edge with weight $\pm h_j$ by $V^{(t+1)}_j$ (this is the final variance of the iteration). From claim \ref{product_claim}, we then get:
\begin{eqnarray} \label{var_prod}
\frac{1}{V^{(t+1)}_j} = \sum_{\substack{i=1 \\ i \neq j}}^d \frac{1}{\tilde{V}^{(t)}_i} + \frac{1}{\sigma^2}
\end{eqnarray}
From symmetry considerations, it can be seen that all messages that are sent along edges with the same absolute value of their weight will have the same variance, since the same variance update occurs for all these messages (both for check node messages and variable node messages). Therefore, the $d$ variance values $V^{(t)}_1, V^{(t)}_2,...,V^{(t)}_d$ are the same for all variable nodes, where $V^{(t)}_{l}$ is the variance of the message that is sent along an edge with weight $\pm h_l$. This completes the proof of Lemma \ref{same_val_same_var}.

Using this symmetry, we can now derive the recursive update of the variance values $V^{(t)}_1, V^{(t)}_2,...,V^{(t)}_d$. Substituting (\ref{var_check}) in (\ref{var_prod}), we get:
\begin{eqnarray} \label{var_recur_app_eq}
\frac{1}{V^{(t+1)}_i} = \frac{1}{\sigma^2} + \sum_{\substack{m=1 \\ m \neq i}}^d \frac{h_m^2}{\sum_{\substack{j=1 \\ j \neq m}}^d h_j^2 V^{(t)}_j}
\end{eqnarray}
for $i=1,2,...d$, which completes the proof of Lemma \ref{var_recur_lemma}.

\subsection{Proof of Theorem \ref{var_behavior}}
We would like to analyze the convergence of the nonlinear recursion (\ref{var_recur}) for the variances $V^{(t)}_1, V^{(t)}_2,...,V^{(t)}_d$. This recursion is illustrated in (\ref{var_recur_3}) for the case $d=3$. It is assumed that $\alpha < 1$, where $\alpha = \frac{\sum_{i=2}^d h_i^2}{h_1^2}$. Define another set of variables $U^{(t)}_1, U^{(t)}_2,...,U^{(t)}_d$, which obey the following recursion. The recursion for the first variable is:
\begin{align} \label{var_tilde_recur}
\frac{1}{U^{(t+1)}_1} = \frac{1}{\sigma^2} + \sum_{m=2}^d \frac{h_m^2}{ h_1^2 U^{(t)}_1}
\end{align}
where for $i=2,3,...d$ the recursion is:
\begin{align*}
\frac{1}{U^{(t+1)}_i} = \frac{h_1^2}{\sum_{j=2}^d h_j^2 U^{(t)}_j}
\end{align*}
with initial conditions $U^{(0)}_1 = U^{(0)}_2   =... = U^{(0)}_d = \sigma^2$.

It can be seen that (\ref{var_tilde_recur}) can be regarded as the approximation of (\ref{var_recur}) under the assumptions that $V^{(t)}_i << V^{(t)}_1$ and $V^{(t)}_i << \sigma^2$ for $i=2,3,...d$. 

For illustration, the new recursion for the case $d=3$ is:
\begin{eqnarray} \label{var_tilde_recur_3}
\frac{1}{U^{(t+1)}_1} = \frac{h_2^2}{h_1^2 U^{(t)}_1} + \frac{h_3^2}{h_1^2 U^{(t)}_1} + \frac{1}{\sigma^2}
\end{eqnarray}
\begin{displaymath}
\frac{1}{U^{(t+1)}_2} = \frac{h_1^2}{h_2^2 U^{(t)}_2 + h_3^2 U^{(t)}_3} \end{displaymath}
\begin{displaymath}
\frac{1}{U^{(t+1)}_3} = \frac{h_1^2}{h_2^2 U^{(t)}_2 + h_3^2 U^{(t)}_3} \end{displaymath}

It can be seen that in the new recursion, $U^{(t)}_1$ obeys a recursion that is independent of the other variables. From (\ref{var_tilde_recur}), this recursion can be written as
$\frac{1}{U^{(t+1)}_1} = \frac{1}{\sigma^2} + \frac{\alpha}{U^{(t)}_1}$, with initial condition $U^{(0)}_1 = \sigma^2$. Since $\alpha < 1$, this is a stable linear recursion for $\frac{1}{U^{(t)}_1}$, which can be solved to get $U^{(t)}_1=\sigma^2(1-\alpha) \frac{1}{1-\alpha^{t+1}}$.

For the other variables, it can be seen that all have the same right hand side in the recursion (\ref{var_tilde_recur}). Since all are initialized with the same value, it follows that $U^{(t)}_2 = U^{(t)}_3   =... = U^{(t)}_d$ for all $t \geq 0$. Substituting back in (\ref{var_tilde_recur}), we get the recursion $U^{(t+1)}_2 = \alpha U^{(t)}_2$, with initial condition $U^{(0)}_2 = \sigma^2$. Since $\alpha < 1$, this is a stable linear recursion for $U^{(t)}_2$, which can be solved to get $U^{(t)}_2 = \sigma^2 \alpha^t$.

We found an analytic solution for the variables $U^{(t)}_i$. However, we are interested in the variances $V^{(t)}_i$. The following claim relates the two sets of variables.

\begin{claim} \label{var_tilde_claim}
For every $t \geq 0$, the first variables of the two sets are related by $V^{(t)}_1 \geq U^{(t)}_1$, where for $i=2,3,...d$ we have $V^{(t)}_i \leq U^{(t)}_i$.
\end{claim}

\begin{proof}
By induction: the initialization of the two sets of variables obviously satisfies the required relations. Assume now that the relations are satisfied for iteration $t$, i.e. $V^{(t)}_1 \geq U^{(t)}_1$ and for $i=2,3,...d$, $V^{(t)}_i \leq U^{(t)}_i$. If we now compare the right hand side of the update recursion for $\frac{1}{V^{(t+1)}_1}$ to that of $\frac{1}{U^{(t+1)}_1}$ (i.e. (\ref{var_recur}) to (\ref{var_tilde_recur})), then the right hand side for $\frac{1}{V^{(t+1)}_1}$ is smaller, because it has additional positive terms in the denominators, where the common terms in the denominators are larger according to the induction assumption. Therefore, $V^{(t+1)}_1 \geq U^{(t+1)}_1$, as required. In the same manner, if we compare the right hand side of the update recursion for $\frac{1}{V^{(t+1)}_i}$ to that of $\frac{1}{U^{(t+1)}_i}$ for $i \geq 2$, then the right hand side for $\frac{1}{V^{(t+1)}_i}$ is larger, because it has additional positive terms, where the common terms are also larger since their denominators are smaller due to the induction assumption. Therefore, $V^{(t+1)}_i \leq U^{(t+1)}_i$ for $i=2,3,...d$, as required.
\end{proof}

Using claim \ref{var_tilde_claim} and the analytic results for $U^{(t)}_i$, we now have:
\begin{align}
V^{(t)}_1 \geq U^{(t)}_1 = \sigma^2(1-\alpha) \frac{1}{1-\alpha^{t+1}} \geq \sigma^2(1-\alpha)
\end{align}
where for $i=2,3,...d$ we have:
\begin{align}
V^{(t)}_i \leq U^{(t)}_i = \sigma^2 \alpha^t
\end{align}
We have shown that the first variance is lower bounded by a positive nonzero constant where the other variances are upper bounded by a term that decays exponentially to zero. Therefore, for large $t$ we have $V^{(t)}_i << V^{(t)}_1$ and $V^{(t)}_i << \sigma^2$ for $i=2,3,...d$. It then follows that for large $t$ the variances approximately obey the recursion (\ref{var_tilde_recur}), which was built from the actual variance recursion (\ref{var_recur}) under these assumptions. Therefore, for $i=2,3,...d$ the variances are not only upper bounded by an exponentially decaying term, but actually approach such a term, where the first variance actually approaches the constant $\sigma^2 (1-\alpha)$ in an exponential rate. This completes the proof of Theorem \ref{var_behavior}.

Note that the above analysis only applies if $\alpha < 1$. To illustrate the behavior for $\alpha \geq 1$, consider the simple case of $h_1=h_2=...=h_d$. From (\ref{var_recur}), (\ref{var_recur_3}) it can be seen that for this case, if $V^{(0)}_i$ is independent of $i$, then $V^{(t)}_i$ is independent of $i$ for every $t>0$, since all the elements will follow the same recursive equations. Substituting this result in the first equation, we get the single variable recursion
$\frac{1}{V^{(t+1)}_i} = \frac{1}{V^{(t)}_i} + \frac{1}{\sigma^2}$
with initialization
$V^{(0)}_i = \sigma^2$.
This recursion is easily solved to get
$\frac{1}{V^{(t)}_i} = \frac{t+1}{\sigma^2}$
or
$V^{(t)}_i = \frac{\sigma^2}{t+1}$.
It can be seen that all the variances converge to zero, but with slow convergence rate of $o(1/t)$.

\section{Asymptotic Behavior of the Mean Values Recursion}
\label{mean_app}
\subsection{Proof of Lemma \ref{eq_mean_lemma} and Lemma \ref{narrow_mean_lemma} (Mean of Narrow Messages)}
Assume a magic square LDLC with dimension $n$ and degree $d$. We shall now examine the effect of the calculations in the check nodes and variable nodes on the mean values and derive the resulting recursion.
Every iteration, each check node generates $d$ output messages, one for each variable node that connects to it, where the weights of these $d$ connections are $\pm h_1, \pm h_2,...,\pm h_d$. For each such output message, the check node convolves $d-1$ expanded variable node PDF messages, and then stretches and periodically extends the result. We shall concentrate on the $nd$ consistent Gaussians that relate to the same integer vector $\underline{b}$ (one Gaussian in each message), and analyze them jointly. For convenience, we shall refer to the mean value of the relevant consistent Gaussian as the mean of the message.

Consider now a specific check node. Denote the mean value of the variable node message that arrives at iteration $t$ along the edge with weight $\pm h_j$ by $m^{(t)}_j$, $j=1,2,...d$. Denote the mean value of the message that is sent back to a variable node along an edge with weight $\pm h_j$ by $\tilde{m}^{(t)}_j$. From (\ref{iter_conv}), (\ref{check_message}) and claim \ref{conv_claim}, we get:
\begin{eqnarray} \label{m_check}
\tilde{m}^{(t)}_j = \frac{1}{h_j}\left(b_k - \sum_{\substack{i=1 \\ i \neq j}}^d h_i m^{(t)}_i\right)
\end{eqnarray}
where $b_k$ is the appropriate element of $\underline{b}$ that is related to this specific check equation, which is the only relevant index in the infinite sum of the periodic extension step (\ref{check_message}). Note that the check node operation is equivalent to extracting the value of $m_j$ from the check equation $\sum_{i=1}^d h_i m_i = b_k$, assuming all the other $m_i$ are known. Note also that the coefficients $h_j$ should have a random sign. To keep notations simple, we assume that $h_j$ already includes the random sign. Later, when several equations will be combined together, we should take it into account.

Then, each variable node generates $d$ messages, one for each check node that is connected to it, where the weights of these $d$ connections are $\pm h_1, \pm h_2,...,\pm h_d$. For each such output message, the variable node generates the product of $d-1$ check node messages and the channel noise PDF. For a specific variable node, denote the mean value of the message that arrives from a check node along an edge with weight $\pm h_j$ by $\tilde{m}^{(t)}_j$, and the appropriate variance by $\tilde{V}^{(t)}_j$. The mean value of the message that is sent back to a check node along an edge with weight $\pm h_j$ is $m^{(t+1)}_j$, the final mean value of the iteration. From claim \ref{product_claim}, we then get:
\begin{eqnarray} \label{mean_prod}
m^{(t+1)}_j = \frac{y_k/\sigma^2 + \sum_{\substack{i=1 \\ i \neq j}}^d \tilde{m}^{(t)}_i /\tilde{V}^{(t)}_i}{1/\sigma^2 + \sum_{\substack{i=1 \\ i \neq j}}^d 1/\tilde{V}^{(t)}_i}
\end{eqnarray}
where $y_k$ is the channel observation for the variable node and $\sigma^2$ is the noise variance. Note that $\tilde{m}^{(t)}_i$, $i=1,2,...,d$ in (\ref{mean_prod}) are the mean values of check node messages that arrive to the same variable node from different check nodes, where in (\ref{m_check}) they define the mean values of check node messages that leave the same check node. However, it is beneficial to keep the notations simple, and we shall take special care when (\ref{mean_prod}) and (\ref{m_check}) are combined.

It can be seen that the convergence of the mean values is coupled to the convergence of the variances (unlike the recursion of the variances which was autonomous). However, as the iterations go on, this coupling disappears. To see that, recall from Theorem \ref{var_behavior} that for each check node, the variance of the variable node message that comes along an edge with weight $\pm h_1$ approaches a finite value, where the variance of all the other messages approaches zero exponentially. According to (\ref{var_check}), the variance of the check node message is a weighted sum of the variances of the incoming variable node messages. Therefore, the variance of the check node message that goes along an edge with weight $\pm h_1$ will approach zero, since the weighted sum involves only zero-approaching variances. All the other messages will have finite variance, since the weighted sum involves the non zero-approaching variance. To summarize, each variable node sends (and each check node receives) $d-1$ ``narrow'' messages and a single ``wide'' message. Each check node sends (and each variable node receives) $d-1$ ``wide'' messages and a single ``narrow'' message, where the narrow message is sent along the edge from which the wide message was received (the edge with weight $\pm h_1$).

We shall now concentrate on the case where the variable node generates a narrow message. Then, the sum in the nominator of (\ref{mean_prod}) has a single term for which $\tilde{V}^{(t)}_i \rightarrow 0$, which corresponds to $i=1$. The same is true for the sum in the denominator. Therefore, for large $t$, all the other terms will become negligible and we get:
\begin{eqnarray} \label{mean_app_prod}
m^{(t+1)}_j \approx \tilde{m}^{(t)}_1
\end{eqnarray}
where $\tilde{m}^{(t)}_1$ is the mean of the message that comes from the edge with weight $h_1$, i.e. the narrow check node message. As discussed above, $d-1$ of the $d$ variable node messages that leave the same variable node are narrow. From (\ref{mean_app_prod}) it comes out that for large $t$, all these $d-1$ narrow messages will have the same mean value. This completes the proof of Lemma \ref{eq_mean_lemma}.

Now, combining (\ref{m_check}) and (\ref{mean_app_prod}) (where the indices are arranged again, as discussed above), we get:
\begin{eqnarray} \label{mean_app_tot}
m^{(t+1)}_{l_1} \approx \frac{1}{h_1}\left(b_k - \sum_{i=2}^d h_i m^{(t)}_{l_i}\right)
\end{eqnarray}
where $l_i$, $i=1,2...,d$ are the variable nodes that take place in the check equation for which variable node $l_1$ appears with coefficient $\pm h_1$. $b_k$ is the element of $\underline{b}$ that is related to this check equation. $m^{(t+1)}_{l_1}$ denotes the mean value of the $d-1$ narrow messages that leave variable node $l_1$ at iteration $t+1$. $m^{(t)}_{l_i}$ is the mean value of the narrow messages that were generated at variable node $l_i$ at iteration $t$. Only narrow messages are involved in (\ref{mean_app_tot}), because the right hand side of (\ref{mean_app_prod}) is the mean value of the \emph{narrow} check node message that arrived to variable node $l_1$, which results from the convolution of $d-1$ \emph{narrow} variable node messages.
Therefore, for large $t$, the mean values of the narrow messages are decoupled from the mean values of the wide messages (and also from the variances), and they obey an autonomous recursion.

The mean values of the narrow messages at iteration $t$ can be arranged in an $n$-element column vector $\underline{m}^{(t)}$ (one mean value for each variable node).
We would like to show that the mean values converge to the coordinates of the lattice point $\underline{x} = \boldsymbol{G} \underline{b}$. Therefore, it is useful to define the error vector $\underline{e}^{(t)} \stackrel{\Delta}{=} \underline{m}^{(t)} - \underline{x}$. Since $\boldsymbol{H} \underline{x} = \underline{b}$, we can write (using the same notations as (\ref{mean_app_tot})):
\begin{eqnarray} \label{x_app_tot}
x_{l_1} = \frac{1}{h_1}\left(b_k - \sum_{i=2}^d h_i x_{l_i}\right)
\end{eqnarray}
Subtracting (\ref{x_app_tot}) from (\ref{mean_app_tot}), we get:
\begin{eqnarray} \label{e_app_tot}
e^{(t+1)}_{l_1} \approx -\frac{1}{h_1}\sum_{i=2}^d h_i e^{(t)}_{l_i}
\end{eqnarray}
Or, in vector notation:

\begin{eqnarray}  \label{thin_mean_recur_app}
\underline{e}^{(t+1)} \approx -\tilde{\boldsymbol{H}} \cdot \underline{e}^{(t)}
\end{eqnarray}
where $\tilde{\boldsymbol{H}}$ is derived from $\boldsymbol{H}$ by permuting the rows such that the $\pm h_1$ elements will be placed on the diagonal, dividing each row by the appropriate diagonal element ($h_1$ or $-h_1$), and then nullifying the diagonal.
Note that in order to simplify the notations, we embedded the sign of $\pm h_j$ in $h_j$ and did not write it implicitly. However, the definition of $\tilde{\boldsymbol{H}}$ solves this ambiguity. This completes the proof of Lemma \ref{narrow_mean_lemma}.

\subsection{Proof of Lemma \ref{wide_mean_lemma} (Mean of Wide Messages)}
Recall that each check node receives $d-1$ narrow messages and a single wide message. The wide message comes along the edge with weight $\pm h_1$. Denote the appropriate lattice point by $\underline{x}=\boldsymbol{G}\underline{b}$, and assume that the Gaussians of the narrow variable node messages have already converged to impulses at the corresponding lattice point coordinates (Theorem \ref{narrow_mean_theorem}). We can then substitute in (\ref{m_check}) $m^{(t)}_i = x_i$ for $i \geq 2$. The mean value of the (wide) message that is returned along the edge with weight $\pm h_j$ ($j \neq 1$) is:
\begin{align} \label{wide_m_check}
\tilde{m}^{(t)}_j = \frac{1}{h_j}\left(b_k - \sum_{\substack{i=2 \\ i \neq j}}^d h_i x_i - h_1 m^{(t)}_1\right) =
\end{align}
\begin{align*}
%\nonumber\\
= \frac{1}{h_j} \left(h_1 x_1 + h_j x_j - h_1 m^{(t)}_1\right) = x_j + \frac{h_1}{h_j}\left(x_1 - m^{(t)}_1\right)
\end{align*}
As in the previous section, for convenience of notations we embed the sign of $\pm h_j$ in $h_j$ itself. The sign ambiguity will be resolved later.

The meaning of (\ref{wide_m_check}) is that the returned mean value is the desired lattice coordinate plus an error term that is proportional to the error in the incoming wide message. From (\ref{var_check}), assuming that the variance of the incoming wide message has already converged to its steady state value $\sigma^2 (1-\alpha)$ and the variance of the incoming narrow messages has already converged to zero, the variance of this check node message will be:
\begin{eqnarray} \label{wide_var_check}
\tilde{V}^{(t)}_j = \frac{h_1^2}{h_j^2} \sigma^2 (1-\alpha)
\end{eqnarray}
where $\alpha = \frac{\sum_{i=2}^d h_i^2}{h_1^2}$. Now, each variable node receives $d-1$ wide messages and a single narrow message. The mean values of the wide messages are according to (\ref{wide_m_check}) and the variances are according to (\ref{wide_var_check}). The single wide message that this variable node generates results from the $d-1$ input wide messages and it is sent along the edge with weight $\pm h_1$. From (\ref{mean_prod}), the wide mean value generated at variable node $k$ will then be:
\begin{align} \label{wide_mean}
m^{(t+1)}_k =
\end{align}
\begin{align*}
 = \frac{y_k/\sigma^2 + \sum_{j=2}^d \left(x_k + \frac{h_1}{h_j}(x_{p(k,j)} - m^{(t)}_{p(k,j)})\right)\frac{h_j^2}{h_1^2 \sigma^2 (1-\alpha)} }{1/\sigma^2 + \sum_{j=2}^d \frac{h_j^2}{h_1^2 \sigma^2 (1-\alpha)}}
\end{align*}
Note that the $x_1$ and $m_1$ terms of (\ref{wide_m_check}) were replaced by $x_{p(k,j)}$ and $m_{p(k,j)}$, respectively, since for convenience of notations we denoted by $m_1$ the mean of the message that came to a check node along the edge with weight $\pm h_1$. For substitution in (\ref{mean_prod}) we need to know the exact variable node index that this edge came from. %, so for the check node that is connected to variable node $k$ with weight $h_j$ we denote it by $l_j$. In other words,
Therefore, $p(k,j)$ denotes the index of the variable node that takes place with coefficient $\pm h_1$ in the check equation where $x_k$ takes place with coefficient $\pm h_j$.

Rearranging terms, we then get:
\begin{align} \label{wide_mean_cont}
m^{(t+1)}_k =
\end{align}
\begin{align*}
 = \frac{y_k (1 - \alpha) + x_k \cdot \alpha + \sum_{j=2}^d \frac{h_j}{h_1}\left(x_{p(k,j)} - m^{(t)}_{p(k,j)}\right)}{(1-\alpha) + \alpha} =
%\nonumber\\
\end{align*}
\begin{align*}
= y_k + \alpha(x_k - y_k) + \frac{1}{h_1} \sum_{j=2}^d h_j (x_{p(k,j)} - m^{(t)}_{p(k,j)})
\end{align*}
Denote now the wide message mean value error by $e^{(t)}_k \stackrel{\Delta}{=} m^{(t)}_k - x_k$ (where $\underline{x}=\boldsymbol{G} \underline{b}$ is the lattice point that corresponds to $\underline{b}$). Denote by $\underline{q}$ the difference vector between $\underline{x}$ and the noisy observation $\underline{y}$, i.e. $\underline{q} \stackrel{\Delta}{=} \underline{y} - \underline{x}$. Note that if $\underline{b}$ corresponds to the correct lattice point that was transmitted, $\underline{q}$ equals the channel noise vector $\underline{w}$. Subtracting $x_k$ from both sides of (\ref{wide_mean_cont}), we finally get:
\begin{eqnarray}
e^{(t+1)}_k = q_k (1-\alpha) - \frac{1}{h_1} \sum_{j=2}^d h_j e^{(t)}_{p(k,j)}
\end{eqnarray}
If we now arrange all the errors in a single column vector $\underline{e}$, we can write:
\begin{eqnarray} \label{wide_mean_recur_app}
\underline{e}^{(t+1)} = -\boldsymbol{F} \cdot \underline{e}^{(t)} + (1-\alpha) \underline{q}
\end{eqnarray}
where $\boldsymbol{F}$ is an $n \times n$ matrix defined by:
\begin{align}
F_{k,l} = \left\{ \begin{array}{ll}
\frac{H_{r,k}}{H_{r,l}} & \textrm{if } k \neq l \textrm{ and there exist a row } r \textrm{ of H} \\
& \textrm{for which } |H_{r,l}|=h_1 \textrm{ and } H_{r,k} \neq 0 \\
0 & \textrm{otherwise}
\end{array} \right.
\end{align}
$\boldsymbol{F}$ is well defined, since for a given $l$ there can be at most a single row of $\boldsymbol{H}$ for which $|H_{r,l}|=h_1$ (note that $\alpha <1$ implies that $h_1$ is different from all the other elements of the generating sequence).

As discussed above, we embedded the sign in $h_i$ for convenience of notations, but when several equations are combined the correct signs should be used. It can be seen that using the notations of (\ref{wide_mean_recur_app}) resolves the correct signs of the $h_i$ elements. This completes the proof of Lemma \ref{wide_mean_lemma}.

An alternative way to construct $\boldsymbol{F}$ from $\boldsymbol{H}$ is as follows. To construct the $k$'th row of $\boldsymbol{F}$, denote by $r_i$, $i=1,2,...d$, the index of the element in the $k$'th column of $\boldsymbol{H}$ with value $h_i$ (i.e. $|H_{r_i, k}| = h_i$). Denote by $l_i$, $i=1,2,...d$, the index of the element in the $r_i$'th row of $\boldsymbol{H}$ with value $h_1$ (i.e. $|H_{r_i, l_i}| = h_1$). The $k$'th row of $\boldsymbol{F}$ will be all zeros except for the $d-1$ elements $l_i$, $i=2,3...d$, where $F_{k, l_i} = \frac{H_{r_i, k}}{H_{r_i, l_i}}$.

\section{Asymptotic Behavior of the Amplitudes Recursion} \label{amp_rec_app}
\subsection{Proof of Lemma \ref{excitation_behavior}}
%\begin{enumerate}
%\item
From (\ref{excitation}), $a^{(t)}_i$ is clearly non-negative. From Sections \ref{var_convergence}, \ref{mean_convergence} (and the appropriate appendices) it comes out that for consistent Gaussians, the mean values and variances of the messages have a finite bounded value and converge to a finite steady state value. The excitation term $a^{(t)}_i$ depends on these mean values and variances according to (\ref{excitation}), so it is also finite and bounded, and it converges to a steady state value, where caution should be taken for the case of a zero approaching variance. Note that at most a single variance in (\ref{excitation}) may approach zero (as explained in Section \ref{var_convergence}, a single narrow check node message is used for the generation of narrow variable node messages, and only wide check node messages are used for the generation of wide variable node messages). The zero approaching variance corresponds to the message that arrives along an edge with weight $\pm h_1$, so assume that $\tilde{V}^{(t)}_{k,1}$ approaches zero and all other variances approach a non-zero value. Then, $\hat{V}^{(t)}_{k,i}$ also approaches zero and we have to show that the term $\frac{\hat{V}^{(t)}_{k,i}}{\tilde{V}^{(t)}_{k,1}}$, which is a quotient of zero approaching terms, approaches a finite value. Substituting for $\hat{V}^{(t)}_{k,i}$, we get:
\begin{align*}
\lim_{\tilde{V}^{(t)}_{k,1} \rightarrow 0} \frac{\hat{V}^{(t)}_{k,i}}{\tilde{V}^{(t)}_{k,1}} = \lim_{\tilde{V}^{(t)}_{k,1} \rightarrow 0} \frac{1}{\tilde{V}^{(t)}_{k,1}}\left(\frac{1}{\sigma^2} + \sum_{\substack{j=1 \\ j \neq i}}^{d}\frac{1}{\tilde{V}^{(t)}_{k,j}}\right)^{-1} =
\end{align*}
\begin{align} \label{var_ratio_lim}
= \lim_{\tilde{V}^{(t)}_{k,1} \rightarrow 0} \left(\frac{\tilde{V}^{(t)}_{k,1}}{\sigma^2} + 1 + \sum_{\substack{j=2 \\ j \neq i}}^{d}\frac{\tilde{V}^{(t)}_{k,1}}{\tilde{V}^{(t)}_{k,j}}\right)^{-1} = 1
\end{align}
Therefore, $a^{(t)}_i$ converges to a finite steady state value, and has a finite value for every $i$ and $t$. This completes the first part of the proof.

%\item
We would now like to show that $\lim_{t \rightarrow \infty} \sum_{i=1}^{nd} a^{(t)}_i$ can be expressed in the form $\frac{1}{2\sigma^2}(\boldsymbol{G} \underline{b} - \underline{y})^T \boldsymbol{W} (\boldsymbol{G} \underline{b} - \underline{y})$. %We shall start with the simple case of $d=2$. For this case, (\ref{excitation}) reduces to:
%\begin{align} \label{exitation_d_eq_2}
%a^{(t)}_{2(k-1)+i} =
%\frac{\left(\tilde{m}^{(t)}_{k,2-i}-y_k\right)^2}{2\left(\sigma^2 + \tilde{V}^{(t)}_{k,2-i}\right)}
%\end{align}
%Every variable node receives and sends 2 messages: a narrow message and a wide message. For the incoming narrow message, as $t \rightarrow \infty$, the mean value approaches the appropriate coordinate of the lattice point $\underline{x}=G\underline{b}$, and the variance approaches zero. Substituting in (\ref{exitation_d_eq_2}), we get that for the outgoing narrow message, the asymptotic excitation term will be:
%\begin{align} \label{exitation_d_eq_2_narrow}
%\lim_{t \rightarrow \infty} a^{(t)}_{2(k-1)+i} =
%\frac{\left(x_k-y_k\right)^2}{2\sigma^2}
%\end{align}
Every variable node sends $d-1$ narrow messages and a single wide message. We shall start by calculating $a^{(t)}_i$ that corresponds to a narrow message. For this case, $d-1$ check node messages take place in the sums of (\ref{excitation}), from which a single message is narrow and $d-2$ are wide. The narrow message arrives along the edge with weight $\pm h_1$, and has variance $\tilde{V}^{(t)}_{k,1} \rightarrow 0$. Substituting in (\ref{excitation}), and using (\ref{var_ratio_lim}), we get:
\begin{align} \label{narrow_excitation}
a^{(t)}_{(k-1)d+i} \rightarrow
\frac{1}{2} \left( \sum_{\substack{j=2 \\ j \neq i}}^{d} \frac{\left(\tilde{m}^{(t)}_{k,1}-\tilde{m}^{(t)}_{k,j}\right)^2}{\tilde{V}^{(t)}_{k,j}} +  \frac{\left(\tilde{m}^{(t)}_{k,1}-y_k\right)^2}{\sigma^2}\right)
\end{align}
Denote $\underline{x} = \boldsymbol{G} \underline{b}$. The mean values of the narrow check node messages converge to the appropriate lattice point coordinates, i.e. $\tilde{m}^{(t)}_{k,1} \rightarrow x_k$. From Theorem \ref{wide_mean_theorem}, the mean value of the wide variable node message that originates from variable node $k$ converges to $x_k+e_k$, where $\underline{e}$ denotes the vector of error terms. The mean value of a wide check node message that arrives to node $k$ along an edge with weight $\pm h_j$ can be seen to approach $\tilde{m}^{(t)}_{k,j}=x_k-\frac{h_1}{h_j}e_{p(k,j)}$, where $p(k,j)$ denotes the index of the variable node that takes place with coefficient $\pm h_1$ in the check equation where $x_k$ takes place with coefficient $\pm h_j$. For convenience of notations, we shall assume that $h_j$ already includes the sign (this sign ambiguity will be resolved later). The variance of the wide variable node messages converges to $\sigma^2(1-\alpha)$, so the variance of the wide check node message that arrives to node $k$ along an edge with weight $\pm h_j$ can be seen to approach $\tilde{V}^{(t)}_{k,j} \rightarrow \frac{h_1^2}{h_j^2}\sigma^2(1-\alpha)$. Substituting in (\ref{narrow_excitation}), and denoting $\underline{q}=\underline{y}-\underline{x}$, we get:
\begin{align*} \label{final_narrow_excitation}
a^{(t)}_{(k-1)d+i} \rightarrow
\frac{1}{2} \left( \sum_{\substack{j=2 \\ j \neq i}}^{d} \frac{\left(\frac{h_1}{h_j}e_{p(k,j)}\right)^2}{\frac{h_1^2}{h_j^2}\sigma^2(1-\alpha)} +  \frac{\left(x_k-y_k\right)^2}{\sigma^2}\right) =
\end{align*}
\begin{align}
= \frac{1}{2\sigma^2}\left[\left(\frac{1}{1-\alpha}\sum_{\substack{j=2 \\ j \neq i}}^{d} e_{p(k,j)}^2 \right) + q_k^2\right]
\end{align}
Summing over all the narrow messages that leave variable node $k$, we get:
\begin{align} \label{sum_narrow_excitation}
\sum_{i=2}^d a^{(t)}_{(k-1)d+i} \rightarrow
\end{align}
\begin{align*}
\rightarrow \frac{1}{2\sigma^2}\left[\left(\frac{d-2}{1-\alpha}\sum_{j=2}^{d} e_{p(k,j)}^2 \right) + (d-1)q_k^2\right]
\end{align*}
To complete the calculation of the contribution of node $k$ to the excitation term, we still have to calculate $a^{(t)}_i$ that corresponds to a wide message. Substituting $\tilde{m}^{(t)}_{k,j} \rightarrow x_k-\frac{h_1}{h_j}e_{p(k,j)}$, $\tilde{V}^{(t)}_{k,j} \rightarrow \frac{h_1^2}{h_j^2}\sigma^2(1-\alpha)$, $\hat{V}^{(t)}_{k,1} \rightarrow \sigma^2(1-\alpha)$ in (\ref{excitation}), we get:
\begin{align*}
a^{(t)}_{(k-1)d+1} \rightarrow
\frac{1}{2} \sum_{l=2}^{d} \sum_{j=l+1}^{d} \frac{\left(\frac{h_1}{h_l}e_{p(k,l)}-\frac{h_1}{h_j}e_{p(k,j)}\right)^2}{\frac{h_1^2}{h_l^2}\cdot
\frac{h_1^2}{h_j^2}\sigma^2(1-\alpha)} +
\end{align*}
\begin{align} \label{wide_excitation}
+\frac{1}{2}\sum_{l=2}^{d}\frac{\left(x_k-\frac{h_1}{h_l}e_{p(k,l)}-y_k\right)^2}{\frac{h_1^2}{h_l^2}\sigma^2}
\end{align}
Starting with the first term, we have:
\begin{align} \label{first_term_wide_excitation}
 \sum_{l=2}^{d} \sum_{j=l+1}^{d} \frac{\left(\frac{h_1}{h_l}e_{p(k,l)}-\frac{h_1}{h_j}e_{p(k,j)}\right)^2}{\frac{h_1^2}{h_l^2}\cdot
\frac{h_1^2}{h_j^2}} =
\end{align}
\begin{align*}
=  \frac{1}{2}\sum_{l=2}^{d} \sum_{j=2}^{d} \left(\frac{h_j}{h_1}e_{p(k,l)}-\frac{h_l}{h_1}e_{p(k,j)}\right)^2 =
\end{align*}
\begin{align*}
 = \frac{1}{2} \sum_{l=2}^{d} \sum_{j=2}^{d} \left(\frac{h_j^2}{h_1^2}e_{p(k,l)}^2 + \frac{h_l^2}{h_1^2}e_{p(k,j)}^2 -2\frac{h_j}{h_1}\frac{h_l}{h_1}e_{p(k,l)}e_{p(k,j)}\right) =
\end{align*}
\begin{align*}
 = \alpha \sum_{j=2}^{d} e_{p(k,j)}^2 - \left(\sum_{j=2}^{d} \frac{h_j}{h_1} e_{p(k,j)}\right)^2=
\end{align*}
\begin{align*}
 = \alpha \sum_{j=2}^{d} e_{p(k,j)}^2 - \left(\boldsymbol{F}\cdot \underline{e}\right)_k^2
\end{align*}
where $\boldsymbol{F}$ is defined in Theorem \ref{wide_mean_theorem} and $\left(\boldsymbol{F}\cdot\underline{e}\right)_k$ denotes the $k$'th element of the vector $\left(\boldsymbol{F}\cdot\underline{e}\right)$. Note that using $\boldsymbol{F}$ solves the sign ambiguity that results from embedding the sign of $\pm h_j$ in $h_j$ for convenience of notations, as discussed above.
Turning now to the second term of (\ref{wide_excitation}):
\begin{align} \label{second_term_wide_excitation}
\sum_{l=2}^{d}\frac{\left(x_k-\frac{h_1}{h_l}e_{p(k,l)}-y_k\right)^2}{\frac{h_1^2}{h_l^2}}=
\end{align}
\begin{align*}
=\sum_{l=2}^{d} \left( e_{p(k,l)}^2 + \frac{h_l^2}{h_1^2} q_k^2 +2 q_k e_{p(k,l)}\frac{h_l}{h_1} \right) =
\end{align*}
\begin{align*}
=\left(\sum_{l=2}^{d} e_{p(k,l)}^2 \right) + \alpha q_k^2 +2q_k \left(\boldsymbol{F}\cdot \underline{e}\right)_k =%\sum_{l=2}^{d} \frac{h_l}{h_1} e_{p(k,l)}
\end{align*}
\begin{align*}
=\left(\sum_{l=2}^{d} e_{p(k,l)}^2 \right) + \alpha q_k^2 +2q_k [(1-\alpha)q_k-e_k]= \end{align*}
\begin{align*}
=\left(\sum_{l=2}^{d} e_{p(k,l)}^2 \right) + (2 -\alpha) q_k^2 -2q_k e_k
\end{align*}
where we have substituted $\boldsymbol{F}\underline{e} \rightarrow (1-\alpha)\underline{q}-\underline{e}$, as comes out from Lemma \ref{wide_mean_lemma}. Again, using $\boldsymbol{F}$ resolves the sign ambiguity of $h_j$, as discussed above.

Substituting (\ref{first_term_wide_excitation}) and (\ref{second_term_wide_excitation}) back in (\ref{wide_excitation}), summing the result with (\ref{sum_narrow_excitation}), and rearranging terms, the total contribution of variable node $k$ to the asymptotic excitation sum term is:
\begin{align}
\sum_{i=1}^d a^{(t)}_{(k-1)d+i} \rightarrow
\frac{d-1}{2\sigma^2(1-\alpha)} \sum_{j=2}^{d} e_{p(k,j)}^2 +
\end{align}
\begin{align*}
+\frac{d+1-\alpha}{2\sigma^2}q_k^2 -
\frac{1}{2\sigma^2(1-\alpha)}(\boldsymbol{F}\underline{e})_k^2 -
\frac{1}{\sigma^2}q_k e_k
\end{align*}
Summing over all the variable nodes, the total asymptotic excitation sum term is:
\begin{align}
\sum_{i=1}^{nd} a^{(t)}_{i} = \sum_{k=1}^n \sum_{i=1}^d a^{(t)}_{(k-1)d+i} \rightarrow
\frac{(d-1)^2}{2\sigma^2(1-\alpha)} \left\| \underline{e} \right\|^2 +
\end{align}
\begin{align*}
+\frac{d+1-\alpha}{2\sigma^2} \left\| \underline{q} \right\|^2 -
\frac{1}{2\sigma^2(1-\alpha)}\left\| \boldsymbol{F}\underline{e} \right\|^2 -
\frac{1}{\sigma^2}\underline{q}^T \underline{e}
\end{align*}
Substituting $\underline{e}=(1-\alpha)(\boldsymbol{I}+\boldsymbol{F})^{-1}\underline{q}$ (see Theorem \ref{wide_mean_theorem}), we finally get:
\begin{align}
\sum_{i=1}^{nd} a^{(t)}_{i} \rightarrow \frac{1}{2\sigma^2}\underline{q}^T \boldsymbol{W} \underline{q}
\end{align}
where:
\begin{align*}
\boldsymbol{W} \stackrel{\Delta}{=}  (1-\alpha)(\boldsymbol{I}+\boldsymbol{F})^{{-1}^T}\left((d-1)^2 \boldsymbol{I} -\boldsymbol{F}^T \boldsymbol{F}\right)(\boldsymbol{I}+\boldsymbol{F})^{-1} +
\end{align*}
\begin{align}
+ (d+1-\alpha)\boldsymbol{I} %- (1-\alpha)(I+F^{-1})^{{-1}^T}(I+F^{-1})^{-1} +
-2(1-\alpha) (\boldsymbol{I}+\boldsymbol{F})^{-1}
\end{align}
From (\ref{excitation}) it can be seen that $\sum_{i=1}^{nd} a^{(t)}_{i}$ is positive for every nonzero $\underline{q}$. Therefore, $\boldsymbol{W}$ is positive definite. % and can be decomposed as $W=M^T M$. we then get:
%\begin{align}
%\sum_{i=1}^{nd} a^{(t)}_{i} \rightarrow \frac{1}{2\sigma^2}\left\| M \underline{q} \right\|^2 = \frac{1}{2\sigma^2}\left\| M (\underline{y}-G\underline{b}) \right\|^2
%\end{align}
This completes the second part of the proof.

%\item
Since $a^{(t)}_i$ is finite and bounded, there exists $m_a$ such that $|a^{(t)}_i| \leq m_a$ for all $1 \leq i \leq nd$ and $t >0$. We then have:
\begin{align*}
\sum_{j=0}^{\infty} \frac{\sum_{i=1}^{nd} a^{(j)}_i}{(d-1)^{2j+2}} \leq \sum_{j=0}^{\infty} \frac{nd \cdot m_a}{(d-1)^{2j+2}} = \frac{n \cdot m_a}{(d-2)}
\end{align*}
Therefore, for $d>2$ the infinite sum will have a finite steady state value. This completes the proof of Lemma \ref{excitation_behavior}.
%For $d>2$, the excitation terms are exponentially weighted inside the sum (\ref{norm_sum_amp_solution}), so $\tilde{s}^{(t)}$ will also have a steady state value as $t$ approaches infinity.
%\end{enumerate}

\section{Generation of a Parity Check Matrix for LDLC} \label{LDLC_gen_app}
In the following pseudo-code description, the $i,j$ element of a matrix $P$ is denoted by $P_{i,j}$ and the $k$'th column of a matrix $P$ is denoted by $P_{:,k}$.

\begin{tabbing}
\# \emph{Input: }\=\emph{block length $n$, degree $d$,} \\
\> \emph{nonzero elements $\{h_1, h_2,...h_d\}$.} \\
\# \emph{Output: a magic square LDLC parity check matrix $\boldsymbol{H}$} \\
\> \emph{with generating sequence $\{h_1, h_2,...h_d\}$.} \\
\# \emph{Initialization:} \\
\> \emph{choose $d$ random permutations on $\{1,2,...n\}$.} \\
\> \emph{Arrange the permutations in an $d \times n$ matrix $P$ } \\
\> \emph{such that each row holds a permutation.} \\
\\
\> $c = 1$; \=      \# \emph{column index} \\
\\
\> $loopless\_columns = 0$;  \=\# \emph{number of consecutive} \\
\> \>    \# \emph{columns without loops} \\
\\
\# \emph{loop removal:} \\
while \= $loopless\_columns < n$ \\
\> $changed\_permutation = 0$; \\
\> if \=exists $i \neq j$ such that $P_{i,c}=P_{j,c}$ \\
\> \> \# \emph{a 2-loop was found at column c} \\
\> \> $changed\_permutation = i$; \\
\> else \\
\> \>\# \emph{if there is no 2-loop, look for a 4-loop} \\
\> \> if \=exists $c_0 \neq c$ such that $P_{:,c}$ and $P_{:,c_0}$ have \\
\> \> \> two or more common elements \\
\> \> \> \# \emph{a 4-loop was found at column c} \\
\> \> \> $changed\_permutation$ = line of $P$ for which\\
\> \> \> the first common element appears in column $c$; \\
\> \> end \\
\> end \\
\> if $changed\_permutation \neq 0$ \\
\> \> \# \emph{a permutation should be modified to} \\
\> \> \# \emph{remove loop} \\
\> \> choose a random integer $1 \leq i \leq n$; \\
\> \> swap locations $c$ and $i$ in \\
\> \> permutation $changed\_permutation$; \\
\> \> $loopless\_columns = 0$; \\
\> else \\
\> \> \# \emph{no loop was found at column c} \\
\> \> $loopless\_columns = loopless\_columns + 1$; \\
\> end \\
\> \# \emph{increase column index} \\
\> $c = c + 1$; \\
\> if $c > n$ \\
\> \> $c = 1$; \\
\> end \\
end \\
\\
\# \emph{Finally, build H from the permutations} \\
initialize H as an $n \times n$ zero matrix; \\
for $i = 1:n$ \\
\> for $j = 1:d$ \\
\> \> $H_{P_{j,i},i} = h_j \cdot random\_sign$; \\
\> end \\
end \\

\end{tabbing}

\section{Reducing the Complexity of the FFT Calculations} \label{fft_app}
FFT calculation can be made simpler by using the fact that the convolution is followed by the following steps: the convolution result $\tilde{p}_j(x)$ is stretched to $p_j(x) = \tilde{p}_j(-h_j x)$ and then periodically extended to $Q_j(x) =\sum_{i=-\infty}^{\infty}p_j\left(x-\frac{i}{h_j}\right)$ (see (\ref{check_message})). It can be seen that the stretching and periodic extension steps can be exchanged, and the convolution result $\tilde{p}_j(x)$ can be first periodically extended with period $1$ to $\tilde{Q}_j(x) =\sum_{i=-\infty}^{\infty}\tilde{p}_j\left(x+i\right)$ and then stretched to $Q_j(x) = \tilde{Q}_j(-h_j x)$. Now, the infinite sum can be written as a convolution with a sequence of Dirac impulses:
\begin{eqnarray}
\tilde{Q}_j(x) = \sum_{i=-\infty}^{\infty}\tilde{p}_j\left(x+i\right)=
\tilde{p}_j(x) \circledast \sum_{i=-\infty}^{\infty}\delta(x+i)
\end{eqnarray}
Therefore, the Fourier transform of $\tilde{Q}_j(x)$ will equal the Fourier transform of $\tilde{p}_j(x)$ multiplied by the Fourier transform of the impulse sequence, which is itself an impulse sequence. The FFT of $\tilde{Q}_j(x)$ will therefore have several nonzero values, separated by sequences of zeros. These nonzero values will equal the FFT of $\tilde{p}_j(x)$ after decimation. To ensure an integer decimation rate, we should choose the PDF resolution $\Delta$ such that an interval with range $1$ (the period of $\tilde{Q}_j(x)$) will contain an integer number of samples, i.e. $1/\Delta$ should be an integer. Also, we should choose $L$ (the number of samples in $\tilde{Q}_j(x)$) to correspond to a range which equals an integer, i.e. $D = L \cdot \Delta$ should be an integer. Then, we can calculate the (size L) FFT of $\tilde{p}_j(x)$ and then decimate by $D$. The result will give $1/\Delta$ samples which correspond to a single period (with range 1) of $\tilde{Q}_j(x)$.

However, instead of calculating an FFT of length $L$ and immediately decimating, we can directly calculate the decimated FFT. Denote the expanded PDF at the convolution input by $\tilde{f}_k$, $k=1,2,...L$ (where the expanded PDF is zero padded to length $L$). To generate directly the decimated result, we can first calculate the (size $D$) FFT of each group of $D$ samples which are generated by decimating $\tilde{f}_k$ by $L/D=1/\Delta$. Then, the desired decimated result is the FFT (of size $1/\Delta$) of the sequence of first samples of each FFT of size $D$. However, The first sample of an FFT is simply the sum of its inputs. Therefore, we should only calculate the sequence (of length $1/\Delta$) $g_i=\sum_{k=0}^{D-1} \tilde{f}_{i+k/\Delta}$, $i=1,2,...1/\Delta$ and then calculate the FFT (of length $1/\Delta$) of the result. This is done for all the expanded PDF's. Then, $d-1$ such results are multiplied, and an IFFT (of length $1/\Delta$) gives a single period of $\tilde{Q}_j(x)$.

With this method, instead of calculating $d$ FFT's and $d$ IFFT's of size larger than $L$, we calculate $d$ FFT's and $d$ IFFT's of size $L/D=1/\Delta$.
%Note that this method can not be used in the final decision step since the convolution is not followed by periodic extension. However, the final step is performed only once (and not every iteration), so its optimization is less important.

In order to generate the final check node message, we should stretch $\tilde{Q}_j(x)$ to $Q_j(x) = \tilde{Q}_j(-h_j x)$.
This can be done by interpolating a single period of $\tilde{Q}_j(x)$ using interpolation methods similar to those that were used in Section \ref{implementation} for expanding the variable node PDF's.

\section*{Acknowledgment}
% optional entry into table of contents (if used)
%\addcontentsline{toc}{section}{Acknowledgment}
Support and interesting discussions with Ehud Weinstein are gratefully acknowledged.


\begin{thebibliography}{1}

\bibitem{Shannon1}
C.~E.~Shannon, ``A mathematical theory of communication,'' \emph{Bell\ Syst.\
Tech.\ J.}, vol.\ 27, pp.~379-–423 and pp.~623–-656, July and Oct. 1948.

\bibitem{Elias_linear}
P.~Elias, ``Coding for noisy channels,'' in \emph{IRE Conv. Rec.}, Mar. 1955,
vol.\ 3, pt.\ 4, pp.~37-–46.

\bibitem{Shannon2}
C.~E.~Shannon, ``Probability of error for optimal codes in a Gaussian
channel,'' \emph{Bell Syst. Tech. J.}, vol. 38, pp.~611-–656, 1959.

\bibitem{Blahut}
R.~E.~Blahut, \emph{Theory and Practice of Error Control Codes}. Addison-Wesley, 1983.

\bibitem{Gallager}
R.~G.~Gallager, \emph{Low-Density Parity-Check Codes}. Cambridge, MA:
MIT Press, 1963.

\bibitem{turbo}
C.~Berrou, A.~Glavieux, and P.~Thitimajshima, ``Near Shannon limit error-correcting coding and decoding: Turbo codes,'' \emph{Proc. IEEE Int. Conf. Communications}, pp.~1064--1070, 1993.

\bibitem{Debuda1}
R.~de Buda, ``The upper error bound of a new near-optimal code,'' \emph{IEEE
Trans. Inform. Theory}, vol. IT-21, pp.~441-–445, July 1975.

\bibitem{Debuda2}
R.~de Buda, ``Some optimal codes have structure,'' \emph{IEEE J. Select. Areas
Commun.}, vol. 7, pp.~893-–899, Aug. 1989.

\bibitem{correct_Debuda}
T.~Linder, Ch.~Schlegel, and K.~Zeger, ``Corrected proof of de Buda`s
Theorem,'' \emph{IEEE Trans. Inform. Theory}, %vol. 39,
pp.~1735-–1737, Sept. 1993.

\bibitem{Lolieger}
H.~A.~Loeliger, ``Averaging bounds for lattices and linear codes,'' \emph{IEEE
Trans. Inform. Theory}, vol. 43, pp.~1767-–1773, Nov. 1997.

\bibitem{Urbanke_lattice}
R.~Urbanke and B.~Rimoldi, ``Lattice codes can achieve capacity on the
AWGN channel,'' \emph{IEEE Trans. Inform. Theory}, %vol. 44,
pp.~273-–278,
Jan. 1998.

\bibitem{Zamir_Erez}
U.~Erez and R.~Zamir, ``Achieving 1/2 log(1 + SNR) on the AWGN
channel with lattice encoding and decoding,'' \emph{IEEE Trans. Inf. Theory},
vol. 50, pp.~2293-–2314, Oct. 2004.

\bibitem{Calderbank_codes}
A.~R.~Calderbank and N.~J.~A.~Sloane, ``New trellis codes based
on lattices and cosets,'' \emph{IEEE Trans. Inform. Theory}, vol. IT-33, pp.~177-–195, Mar. 1987.

\bibitem{Forney_codes}
G.~D.~Forney, Jr., ``Coset codes—-Part I: Introduction and geometrical
classification,'' \emph{IEEE Trans. Inform. Theory}, %vol. 34,
pp.~1123-–1151,
Sept. 1988.

\bibitem{signal}
O.~Shalvi, N.~Sommer and M.~Feder, ``Signal Codes,'' \emph{proceedings of the Information theory Workshop}, 2003, pp.~332--336.

\bibitem{signal_journal}
O.~Shalvi, N.~Sommer and M.~Feder, ``Signal Codes,'' in preparation.

\bibitem{bennatan_burshtein}
A. Bennatan and D. Burshtein, ``Design and analysis of nonbinary
LDPC codes for arbitrary discrete-memoryless channels,''
\emph{IEEE Transactions on Information Theory}, volume 52, no. 2,
pp.~549--583, February 2006.

\bibitem{hou_et_al}
J. Hou, P.~H Siegel, L.~B Milstein and H.~D Pfister, ``Capacity
approaching bandwidth efficient coded modulation schemes based on
low density parity check codes,'' \emph{IEEE Transactions on
Information Theory}, volume 49, pp. 2141--2155, Sept. 2003.

\bibitem{Sloane}
J.~H.~Conway and N.~J.~Sloane, \emph{Sphere Packings, Lattices and Groups}. New York: Springer, 1988.

\bibitem{Poltyrev}
G.~Poltyrev, ``On coding without restrictions for the AWGN channel,''
\emph{IEEE Trans. Inform. Theory}, vol. 40, pp.~409-–417, Mar. 1994.

%\bibitem{full_paper}
%N.~Sommer, M.~Feder and O.~Shalvi, ``Low Density Lattice Codes,'' in preparation.

\bibitem{Papoulis}
A.~Papoulis, \emph{Probability, Random variables and Stochastic Processes}. McGraw Hill, second edition, 1984.

\bibitem{sparse}
Y.~Saad, \emph{Iterative Methods for Sparse Linear Systems}. Society for Industrial and Applied Mathematic (SIAM), 2nd edition, 2003.

\bibitem{Agrell_lattice}
E.~Agrell, T.~Eriksson, A.~Vardy, and K.~Zeger, ``Closest point search
in lattices,'' \emph{IEEE Trans.\ Inf.\ Theory}, vol.\ 48, pp.~2201-–2214, Aug.\
2002.

\bibitem{Fano_lattice}
N.~Sommer, M.~Feder and O.~Shalvi, ``Closest point search in lattices using sequential decoding,'' \emph{proceedings of the International Symposium on Information Theory (ISIT)}, 2005, pp.~1053--1057 .

\bibitem{LAST}
H.~El~Gamal, G.~Caire and M.~Damen, ``Lattice coding and decoding achieve the optimal diversity-multiplexing tradeoff of MIMO channels,'' \emph{IEEE Trans.\ Inf.\ Theory}, vol.\ 50, pp.~968--985, Jun. 2004.



\end{thebibliography}
\end{document}